\documentclass[11pt]{article}

\usepackage[final]{acl}

\usepackage{times}
\usepackage{latexsym}

\usepackage[T1]{fontenc}

\usepackage[utf8]{inputenc}

\usepackage{microtype}

\usepackage{inconsolata}

\usepackage{graphicx}
\usepackage{enumitem}
\usepackage{multirow}
\usepackage{amssymb}
\usepackage{hyperref}
\usepackage{url}
\usepackage{booktabs}
\usepackage{xcolor}
\usepackage{algorithm}
\usepackage{algpseudocode}
\usepackage{makecell}
\usepackage{array}
\usepackage[most]{tcolorbox}
\newcommand{\ModelName}{{Self-EmoQ}}

\title{Self-EmoQ: Plutchik-Guided Value-based Planning to Drive Streaming Emotional TTS}

\author{
 \textbf{Yue Zhao\textsuperscript{1}},
 \textbf{Hongyan Li\textsuperscript{1}},
 \textbf{Yong Chen\textsuperscript{1}},
 \textbf{Luo Ji\textsuperscript{1}},
\\
\\
 \textsuperscript{1}Geely AI Lab,
\\
 \small{
   \textbf{Correspondence:} \href{Luo.Ji1@geely.com}{ Luo.Ji1@geely.com}
 }
}

\begin{document}

\maketitle
\begin{abstract}
Emotional interaction is increasingly crucial for conversational AI, yet current systems lack a self-emotion determination mechanism to drive the streaming text-to-speech (TTS) synthesis. We propose an emotion-planning framework that determines the emotion prior to the textual generation, grounding the downstream emotional TTS in a streaming manner. The framework is implemented by a plug-and-play LLM module, initialized from pretrained LLMs, and trained by reinforcement learning (RL) with emotions as the actions. A hybrid reward is employed which combines imitation signals with theory-driven scoring, in which the theory of Plutchik's wheel of emotions is adopted. By experiments on DailyDialog, EmoryNLP, IMEOCAP, and MELD, our method outperforms prompting and finetuning baselines on both emotion determination and response quality. We finally implement an entire streaming pipeline for real-time deployment, with the speech quality confirming the framework's emotional alignment, contextual coherence, and expressive fluency. Codes, cases, and demos are available in \url{https://sixingdeguo.github.io/EmoQ-page/}.


\end{abstract}

\section{Introduction}

\begin{figure}[!t]
\centering
  \includegraphics[width=0.95\linewidth]{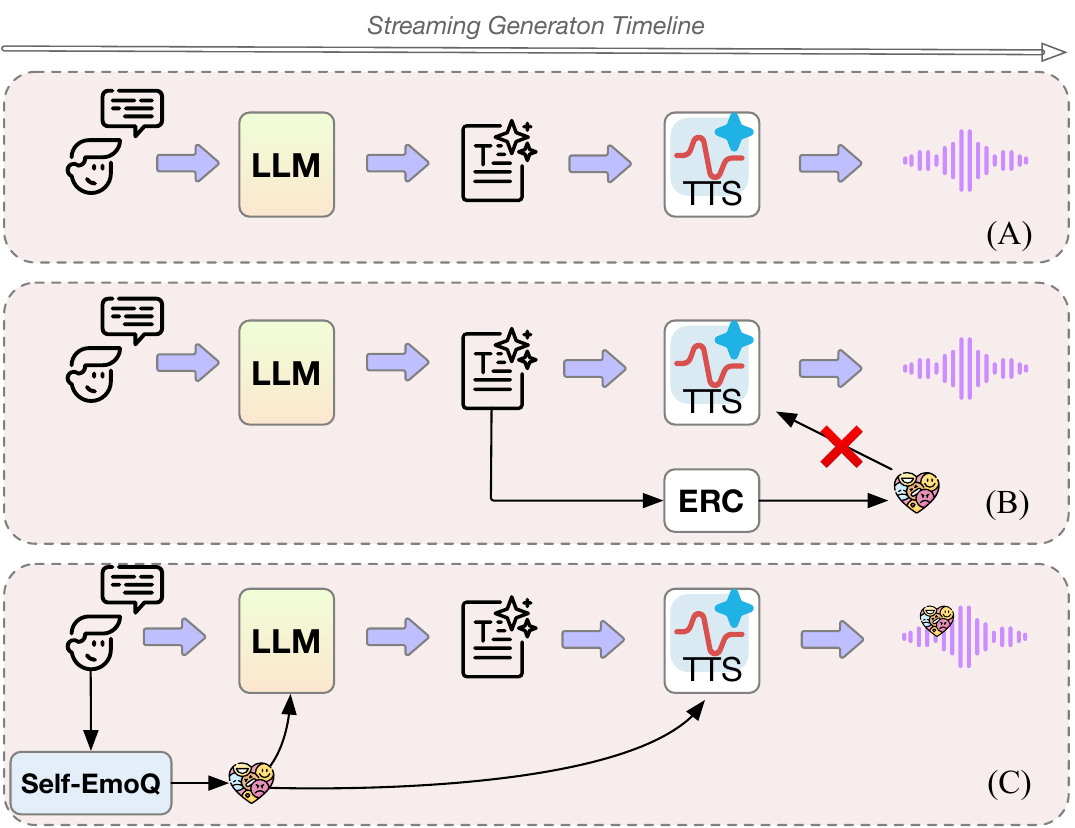} 
  \caption{Comparison of different streaming, emotional LLM-TTS paradigms. (A) Vanilla LLM-TTS pipeline \textcolor{red}{without} $emotion$ consideration. (B) Conventional emotion cognition modules such as ERC can not be directly integrated into streaming Emo-TTS, in which the $emotion$ cognition can \textcolor{red}{not} be yielded until the text generation completes. (C) The plug-and-play {\ModelName} proposed by us determines self-$emotion$ \textbf{prior to} response generation, effectively driving the downstream streaming TTS by emotional conditioning.}
  \label{fig:paradigm}
\end{figure}

Large Language Model (LLM) has revolutionized open-domain and task-oriented dialogue systems, delivering strong semantic understanding, contextual reasoning, and instruction follow-up abilities \citep{lei2023instructerc,yang2023towards,chen2024recent}. In real-time industrial applications, LLM-based conversational agents are usually integrated with Automatic Speech Recognition (ASR) and Text-To-Speech (TTS) modules, forming a cascade pipeline. To improve the response speed, developers have adopted streaming techniques to link text generation with speech synthesis. In this setup, each text token is immediately transformed into the audio segment of TTS (Figure \ref{fig:paradigm}, Setting A).


At the same time, the demand for emotional interaction in conversational AI has grown rapidly. Users not only expect systems to provide accurate information, but also expect responses that convey emotion and empathy. There are substantial studies on text-based emotion studies, including Emotion Recognition in Conversation (ERC) and Emotion Prediction in Conversation (EPC). ERC focuses on identifying the emotional state of the speaker from the speaker's current utterance \citep{poria2017context,majumder2019dialoguernn,ghosal2019dialoguegcn}.
On the other hand, EPC aims to predict the speaker's emotional state in the upcoming turn, based on the knowledge of past utterance and emotion trajectories \citep{shi2024emotional,ju2023real}. On the speech modality, Emotional Text-to-Speech (Emo-TTS) aims to provide stylized speech with controllable prosody, conditioned on predetermined emotion labels \citep{lei2022msemotts,kanda2024making,wu2024laugh}. Together, these developments highlight the feasibility of emotion-aware conversational agents that integrate textual and acoustic affect.


However, these techniques may encounter a critical shortcoming when deployed on industrial streaming implementations. In such a situation, the emotional tone of TTS must be provided at the beginning of generation, while ERC can only recognize the emotion after the entire textual response is completed (Figure \ref{fig:paradigm}, Setting B). In contrast, frameworks that determine the emotion \textbf{before} the response generation can drive the emotional TTS in this streaming manner, as visualized by Figure \ref{fig:paradigm} (Setting C). Such frameworks may be implemented by prompt-based methods which encourage the LLM to decode the emotion first \citep{li2024enhancingemotionalgenerationcapability,10.1007/978-981-95-7078-2_23}, or the EPC method mentioned above. Nevertheless, prompting methods may have suboptimal planning without the model parameter update; while EPC's prediction is fundamentally constrained by supervision, lacking the ability to plan future emotional states and optimize the entire conversation quality, beyond merely imitating dataset trajectories. As a result, we suppose to treat emotion not only as a target to be recognized or predicted, but also as a controllable decision variable that can be explicitly planned over dialogue turns. In this scenario, the reward becomes critical, which should not only represent the ground-truth annotations in the datasets, but also represent human behaviors and generalize to versatile situations.


\textbf{Plutchik’s Wheel of Emotion} \citep{plutchik1982psychoevolutionary} provides a psychological theory on structured emotion categories, intensities, and transition patterns within human interaction. Rather than treating emotions as static labels, this theory reveals regularities in emotional evolution and their functional roles in guiding behavior. Inspired by this theory, we design a relative reward mechanism, to bridge the gap. Correspondingly, we formulate the emotional dialogue generation as a sequential decision-making problem and solve it using reinforcement learning (RL), to obtain an emotional planning module by optimizing the long-term returns. After deploying, this module determines the emotion as its action, which subsequently guides both text generation and emotional speech synthesis (Figure \ref{fig:paradigm}, Setting C).


In this paper, we propose a novel emotional dialogue framework called \textbf{\ModelName}, to determine the self-emotions by bootstrapping the Q-values of value-based RL. Initialized from a pretrained LLM, a plug-and-play planner is trained by the paradigm of Deep Q-Network (DQN) \citep{mnih2015human}, as the upstream module of the LLM generator and the Emo-TTS. We define the dialogue context as the state, while the system’s emotion as the action, forming an utterance-level MDP. The total reward is defined as the linear combination of the \textbf{imitating reward}, determined by the ground-truth of emotion-annotated datasets, and the \textbf{Plutchik score}, annotated by GPT-4o based on the principles proposed by the Plutchik theory. With the Q-values calculated from the average of the output token logprobs, the module is then trained by the Bellman Equation, similar to \citet{wang-etal-2025-convert}. By these mechanisms, we enable long-term emotional planning rather than reactive emotion assignment, and also mimic human interaction patterns by optimizing relative rewards. This module is finally integrated with the streaming LLM-TTS pipeline, providing the emotion conditions that guide the subsequent textual and speech generations. We conduct experiments on DailyDialog, EmoryNLP, IMEOCAP, and MELD, and show that {\ModelName} outperforms prompting, supervised, and tabular Q-learning baselines, on reward optimization, emotional determination accuracy, and qualities of generated response and speech. Our major contributions are summarized as follows:


\begin{itemize}
    \item We propose a novel \textbf{self-emotion planning framework} that integrates value-based RL with LLM-based dialogue generation.
    
    \item We design a theory-driven reward based on \textbf{Plutchik’s Wheel of Emotion}, to align the framework with human emotional behaviors.
    
    \item We implement a \textbf{streaming} pipeline for emotional language and speech generation, and verify its performance on both emotion determination and generation quality.
\end{itemize}

\section{Preliminary}

\paragraph{Utterance-level MDP.}
The Markov decision process (MDP) is usually defined as a 5-tuple $(\mathcal{S}, \mathcal{A}, \mathcal{R}, \mathcal{T}, \gamma)$, where $\mathcal{S}$ is the state set, $\mathcal{A}$ is the action set, $\mathcal{R}$ is the reward set, $\gamma$ is the discounting factor of rewards, and $\mathcal{T}: \mathcal{S} \times \mathcal{A} \rightarrow \mathcal{S}$ is the state transition function.
In this work, we formalize the emotional dialogue task as a strategy-level MDP, with the action space $\mathcal{A} = \{ a \}$ as the set of possible strategies.

\paragraph{Q-Learning.} In value-based RL, the goal is to learn the state-value function $V(s)$ or the state-action value function $Q(s, a)$, such that the determined action achieves the highest expected discounted cumulative reward:
\begin{align}
    a^{\star} = &\arg\max_{a} Q(s, a) \leftarrow \arg\max \sum_{t=0}^{\infty} \gamma^t r(s_t, a_t) \label{eq:maxQ}
\end{align}
which is solved by the famous Bellman Equation: 
\begin{align}
    Q^*(s, a) = r(s,a) + \gamma \max_{a'}Q^*(s', a') \label{eq:bellman} 
\end{align}
in which the superscript $'$ indicates the next step and $r(s,a)$ represents the reward received from environmental interaction. Instead of explicitly implementing the above equation, Deep Q-learning (DQN) approximates the maximization of the right-hand side with the deep value networks:
\begin{align}
    \mathcal{L}(\theta) = | r(s,a) + Q_{\phi}(s', a') - Q_{\theta}(s, a)|^2 \label{eq:dqn}  
\end{align}
where $\mathcal{L}$ is the loss, $\theta$ and $\phi$ are parameters of the Q-net and the target Q-net, respectively. $\phi$ can be periodically synchronized with $\theta$. 

\begin{figure*}[!t]
\centering
  \includegraphics[width=0.95\linewidth]{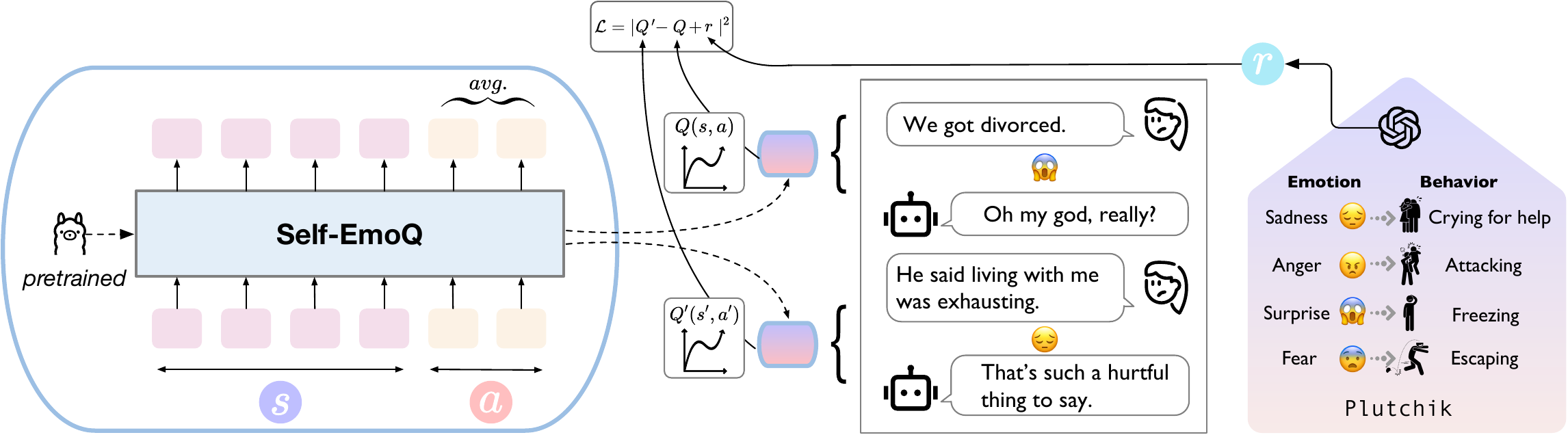} 
  \caption{Framework of {\ModelName}, which is post-trained on pretrained LLM, and produces $Q$-values by averaging output token logprobs. We apply \textit{Plutchik's Wheel of Emotions} to guide reward annotations of multi-turn conversations, and finally update the model based on the Bellman Equation, bootstrapping the long-term emotional return.}
  \label{fig:framework}
\end{figure*}

\section{Methodology}

\subsection{Task definition}

\begin{table}[htbp!]

    \centering
    \small
    
    \begin{tabular}{c|l}
        \toprule
        \textit{Query} & \makecell[l]{\textit{\{history\}} \\
        Joey Tribbiani: (\textit{Neutral}) Hello.\\
        Chandler Bing: (\textit{Mad}) Look I never should \\have kissed your girlfriend, but I'm...}  \\
        \midrule
        \textit{\textcolor{red}{Emotion}} & {\textcolor{red}{Mad} } \\
        \midrule
      \textit{Response} & \makecell[l]{\textit{Joey Tribbiani:} Stop callin'!!}   \\
        \bottomrule
    \end{tabular}
    \caption{An example of \textit{EmoryNLP}.} 
    \label{tab:ex_ESConv}
    \vspace{-2.5mm}
\end{table}

We consider a multi-turn emotional dialogue between a user and an agent. 
At each turn $t$, the user produces an utterance $x_t^u$, and the agent generates a textual response $x_t^s$ (later rendered into speech).  
A conversation session is represented as
\begin{equation}
desc,\; (x_t^u,\, x_t^s)_{t=0:T},
\end{equation}
where $desc$ denotes background information at dialogue level and $T$ is the total number of turns.

To enable controllable emotional generation, the agent additionally selects a \emph{self-emotion} $e_t^s$ at each turn before producing $x_t^s$ and its corresponding speech.
Thus, the sample in turn $t$ is written as
\begin{equation}
(x^u_t,\, e_t^s,\, x_t^s),
\end{equation}
where the dialogue history is defined as
\begin{equation}
h_t = (x_i^u,\, e_i^s,\, x_i^s)_{i=0:t-1}.
\end{equation}

\subsection{Plutchik’s Wheel of Emotion}
\label{sec:theory}

\begin{figure}[!t]
\centering
  \includegraphics[width=0.6\linewidth]{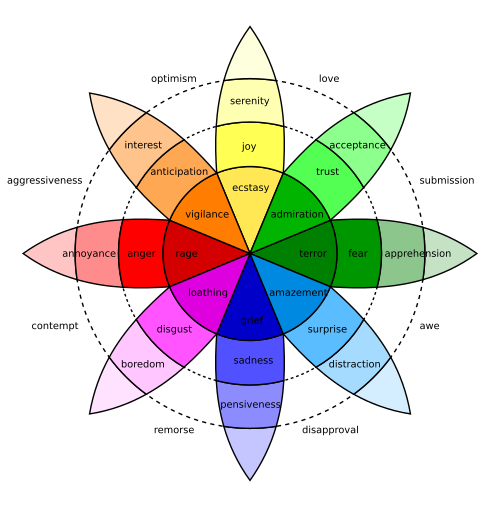} 
  \caption{The Plutchik's wheel of emotions.}
  \label{Plutchik wheel}
  \vspace{-3.5mm}
\end{figure}

Our goal is to enable \emph{natural emotional decision-making} during dialogue. We want to incorporate theoretically grounded principles to guide emotion selection. As the theoretical foundation of our reward design, the theory of \textbf{Plutchik’s Wheel of Emotion} is adopted, which provides a structured theory of emotions, including their categories, opposite and adjacent relationships, and characteristic behavioral functions. Figure \ref{Plutchik wheel} provides the visualization of the emotional taxonomy. This theory enables us to evaluate not only whether a predicted emotion matches a label but also whether it is \emph{reasonable}, \emph{functional}, and \emph{consistent} within the dialogue context. According to the Plutchik theory, the emotional expression of an utterance is closely coupled with its underlying behavioral function. 
Emotional transitions are not arbitrary but tend to follow the topological structure of the emotion wheel, where transitions between adjacent emotions are generally more natural, while transitions between opposite emotions are less plausible.

Motivated by these principles, we design a \textbf{Plutchik Score} $r_{\text{Plu}}(s_t, e_t^s, x_t^s)$ to evaluate the emotional appropriateness of system responses from a theory-driven perspective. GPT-4o is applied to score three dimensions (Emotion Alignment, Emotion Transition Plausibility, and Emotion–Function Consistency) according to Plutchik theory.
The average of three demensions was the Plutchik score $r_{\text{Plu}}(s_t, e_t^s, x_t^s)$. Detailed prompt on the scoring process is in Appendix\ref{Plutchik score}.

\subsection{System Definitions}

We formulate the emotional dialogue system as a Markov Decision Process (MDP).

\begin{equation}
\mathcal{M} = (\mathcal{S},\, \mathcal{A},\, R,\, \mathcal{T},\, \gamma).
\end{equation}

\paragraph{State.}
The state at turn $t$ is the concatenation of history and user's query at turn $t$:
\begin{equation}
s_t = (desc,\; h_t,\; x_t^u) \in \mathcal{S}.\label{eq:state}
\end{equation}

\paragraph{Action.}
The action is the system's emotion:
\begin{equation}
a_t = e_t^s \in \mathcal{A},
\end{equation}
which controls both emotional text generation and emotional speech synthesis.
Then we generate the response $x_t^s = g(s_t,e_s^t)$ by a fixed pretrained LLM$g(\cdot,\cdot)$.

\paragraph{Reward.}
We consider two types of reward: 1) the imitation reward, which is determined from the labeled emotions in training data, and 2) the theoretical reward, which is the \textbf{Plutchik Score} as introduced in Section \ref{sec:theory}. The entire reward is a linear weighted sum of them:
\begin{align}
    r_t(s_t, e_t^s, x_t^s) =  &(1-w) \cdot\mathbf{1}\!\left[e_t^s = \hat{e}_t^s\right] \notag \\
    &+ w \cdot r_{\text{Plu}}(s_t, e_t^s, x_t^s) \label{eq:reward} 
\end{align}
where $\hat{e}_t^s$ is the emotion of the ground truth of the dataset,  $w$ is the weight of the Plutchik Score. 

\paragraph{Transition.}
The transition function $\mathcal{T}$ evolves as follows:
\begin{enumerate}
    \item The history is updated as 
    \begin{equation}
    h_{t+1} = (h_t,\; x_t^u,\; e_t^s,\; x_t^s).\label{eq:history}
    \end{equation}
    \item The user provides the next utterance $x_{t+1}^u$.
\end{enumerate}

The agent selects the emotion according to a policy $\pi(e_t^s \mid s_t)$ and aims to maximize the cumulative discounted reward:
\begin{equation}
\pi^\star = \arg\max_{\pi}\;
\mathbb{E}_{\pi}\!\left[\sum_{t=0}^{T} \gamma^t r(s_t, e_t^s, x_t^s)\right].
\end{equation}

This MDP formulation allows the agent to treat emotion not merely as a descriptive label but as a controllable decision variable that drives coordinated emotional alignment across text and speech modalities.

\subsection{{\ModelName}}

We implement the self-emotion planner as a plug-and-play module, which is initialized from a pretrained LLM parameterized by $\theta$. By finetuning, we repurpose this module to output the state-action value, \textit{i.e.}, choosing the emotion $a_t \in \mathcal{E}_a$ given the state $s_t$. We let $\mathcal{I}(s_t)$ denote an instruction template that encodes the dialogue state, then append the action, yielding the state-action prompt $\mathcal{I}(s_t) \oplus a_t$. Similar to StraQ* \citep{wang-etal-2025-convert}, the LLM estimates the Q-value through the logprobs of the output token:
\begin{equation}
    Q_{\theta}(s_t, a_t) \;\leftarrow\;
    \text{LLM}_{\theta}\big(\mathcal{I}(s_t) \oplus a_t\big),
\end{equation}
where $\leftarrow$ indicates the average of action logits, in which different actions are inferred as the options of the instruction, in a multi-choice question (MCQ) style. We briefly exhibit the instruction $\mathcal{I}(s)$ below: 


\tcbset{
  colframe=black!75!white,
  colback=gray!5!white,
  boxrule=0.5pt,
  arc=2mm,
  left=1mm, right=1mm, top=1mm, bottom=1mm,
  fonttitle=\bfseries,
  before skip=5pt, after skip=5pt
}
\begin{tcolorbox}[title=Prompt Template]
\textbf{Description:} \{$desc$\} \\
\ 
\textbf{History:} \{$h$\} \\   
\textbf{\  User's query:} \{$query$\} \\
\textbf{Please select the most appropriate response emotion from the following options:\\}
\textbf{(1)} \{$Emo_1$\} 
 \textbf{\ (2)} \{$Emo_2$\} $\cdots$ \textbf{\ (K)} \{$Emo_K$\}\\
\textbf{ Please provide your selection in the format of A through G, your selection is:}
\end{tcolorbox}

The training mechanism of {\ModelName} is illustrated by the visualization by Figure \ref{fig:framework} and the pseudo-codes in Algorithm \ref{alg:emoq}.

\subsection{Emotion-Guided Text and Speech Generation}

After finetuning, the module can be deployed into the pipeline, on the upstream of the LLM-based response generator and the Emo-TTS. It determines the optimal emotion for each state by argmax the Q-values:
\begin{equation}
    a_t^\star = \arg\max_{a \in \mathcal{A}} Q_{\theta}(s_t, a).
\end{equation}

The planned emotion is then injected into the response-generation instruction template, guiding the lexical choice, sentiment strength, and stylistic framing.
The LLM then produces an emotionally aligned response $x_t^s$ conditioned on $e_t^s = a_t^\star$.


Finally, the selected self-emotion $e_t^s$ is used to condition an Emoti TTS model. The embedding of emotion modulates prosody, speaking rate, and acoustic style. Since emotion is determined before decoding, the TTS module can operate in a streaming fashion, converting partial text into emotionally coherent speech as it is generated.




\section{Experiment}
\label{sec:experiment}

\subsection{Setting}

\paragraph{Implementation.} Llama3.1-1B-Instruct \citep{llama3modelcard} is employed as the backbone of emotional determination module, while the training-free dialogue generation backbone is Llama3.1-8B-Instruct. Training is conducted with $L=1024$, $\epsilon=0.1$, $C=5$, $B=512$, $lr=1e-5$, $\gamma=0.8$, and the replay buffer size is 50000.

\paragraph{Datasets.} We evaluate {\ModelName} on four widely used conversational emotion datasets: MELD \citep{poria2019meld}, DailyDialog \citep{li2017dailydialog}, EmoryNLP \citep{zahiri2018emotion}, and IEMOCAP \citep{busso2008iemocap}. The statistics of the datasets are shown in Table \ref{tab:dataset_statistics}, and a detailed description can be found in Appendix \ref{introduction dataset}.

\begin{table}[htbp!]
\centering
\resizebox{0.98\columnwidth}{!}{
\begin{tabular}{lccccccc}
\toprule
\multirow{2}{*}{\textbf{Dataset}}  &
\multicolumn{3}{c}{\textbf{\# Conversations}} &
\multicolumn{3}{c}{\textbf{\# Utterances}} &
\multirow{2}{*}{\textbf{\# Emotions}}  \\
\cmidrule(lr){2-4} \cmidrule(lr){5-7}
 & Train & Val & Test & Train & Val & Test &  \\
\midrule
DailyDialog
& 11118 & 1000 & 1000
& 87170 & 8069 & 7740
& 7 \\
EmoryNLP
& 713 & 99 & 85
& 9934 & 1344 & 1328
& 7 \\
MELD
& 1038 & 114 & 280
& 9989 & 1109 & 2610
& 7 \\
IMEOCAP
& 120 & 12 & 31
& 4810 & 1000 & 1523
& 10 \\
\bottomrule
\end{tabular}}
\caption{Statistics of datasets and evaluation metrics.}
\label{tab:dataset_statistics}
\vspace{-3.5mm}
\end{table}

\paragraph{Reward.} We utilized GPT-4o for the generation of $r_\text{Plu}$; the prompts used can be found in Appendix \ref{Plutchik score}. In Appendix \ref{Evaluation of GPT Scoring}, we discuss the reliability of the GPT-based scoring.

\subsection{Metrics}

For emotion determination evaluations, we employ ranking-based metrics including Recall, 
mean reciprocal rank (MRR), and normalized discounted cumulative gain (NDCG). 
We evaluate the quality of emotion decisions by ranking candidate emotions according to the Q-values. 
The resulting rankings are then compared against the corresponding reward signals to compute ranking-based metrics. 

For the generation task, we utilize the BLEU-2 (B-2), Rouge-L (R-L) and Distinct-2 (D-2). 
The first two are similarity-based metrics, while the last encourages response diversity. 
We also conduct human annotations to evaluate the responses. 
We leave the annotation principle and metric details in the Appendix \ref{detail of metrics1} and \ref{detail of metrics2}.

\begin{figure}[htbp!]
    \centering
    \includegraphics[width=0.45\linewidth]{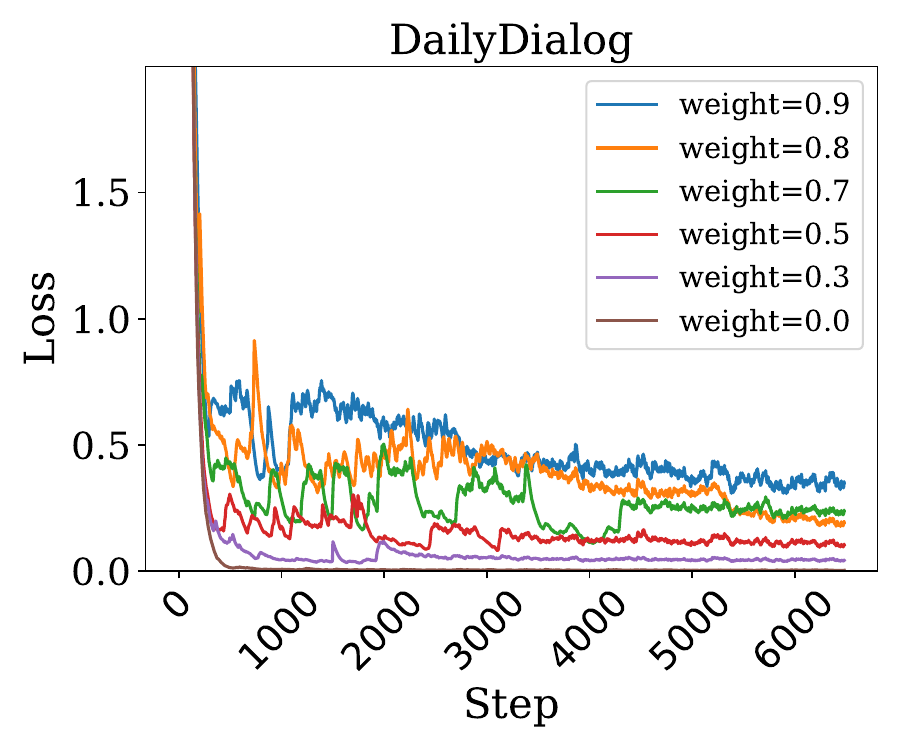}
    \hspace{0.1in}
    \includegraphics[width=0.45\linewidth]{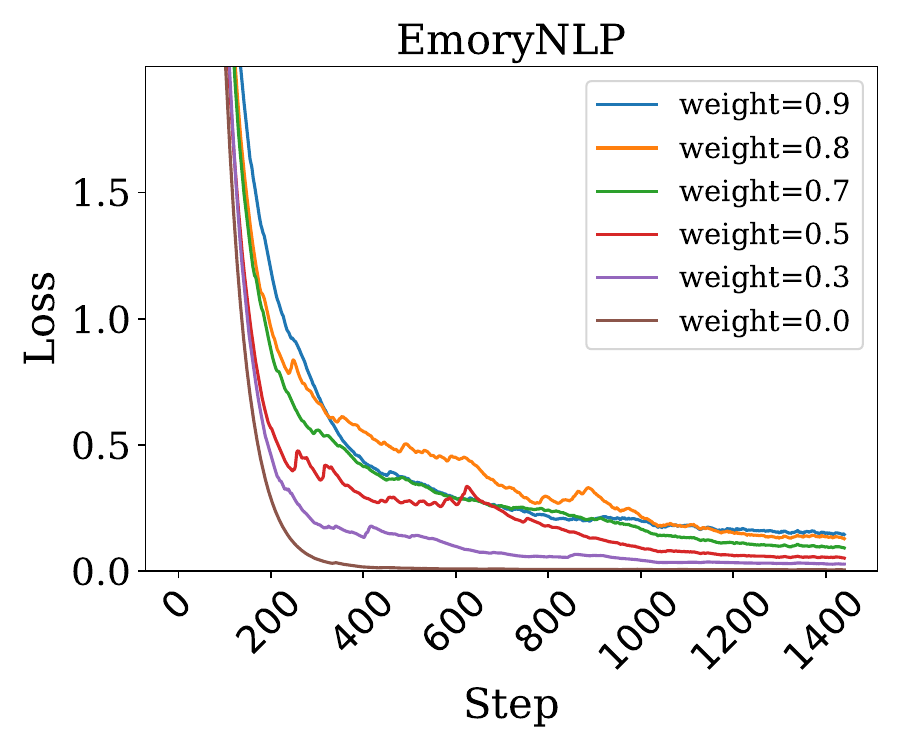}
    \includegraphics[width=0.45\linewidth]{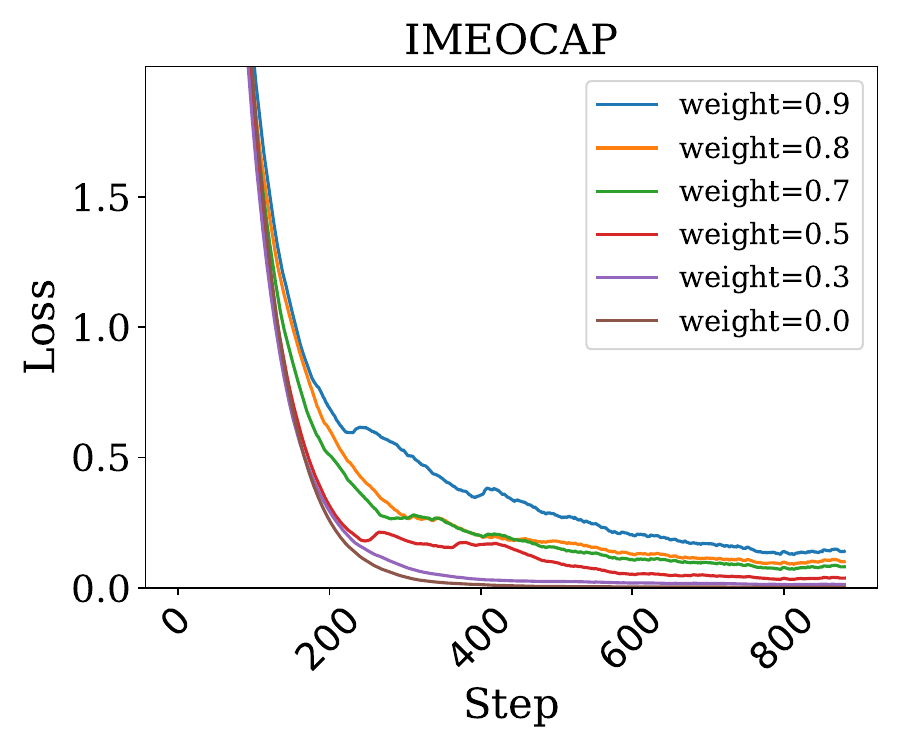}
    \hspace{0.1in}
    \includegraphics[width=0.45\linewidth]{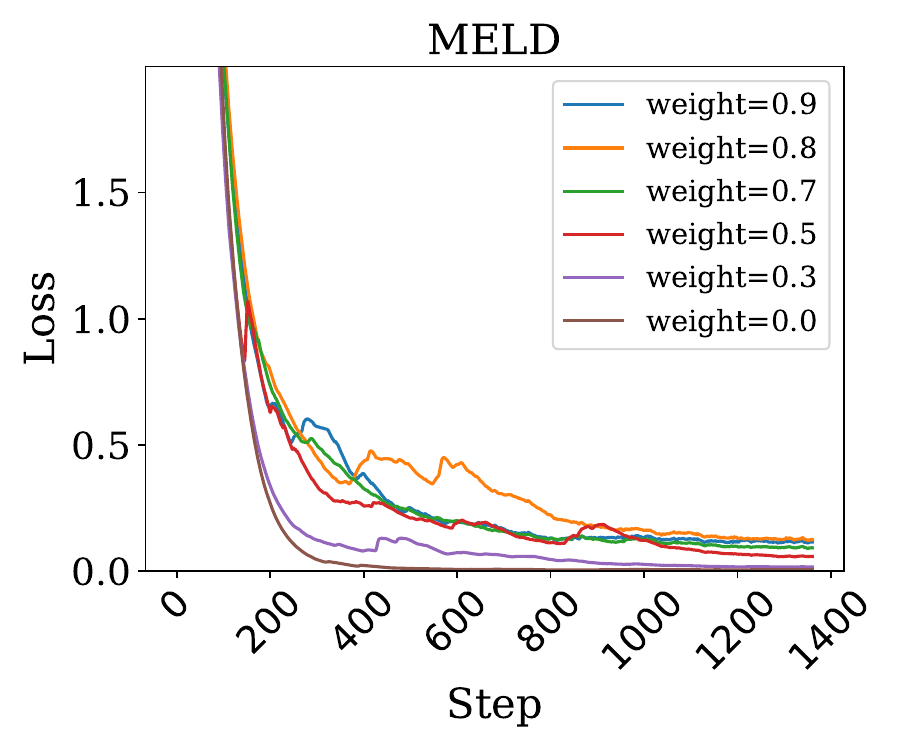}
    \caption{Training loss plots of {\ModelName}.} 
    \label{fig:training_curves}
    \vspace{-3.5mm}
\end{figure}

\begin{table*}[htbp!]
\renewcommand{\arraystretch}{1.1}
\fontsize{14pt}{16pt}\selectfont
\centering
\resizebox{0.98\textwidth}{!}{
\begin{tabular}{p{2.4cm} 
                p{1.0cm}p{1.0cm}p{1.0cm}p{1.0cm}p{1.0cm}p{0.005cm}
                p{1.0cm}p{1.0cm}p{1.0cm}p{1.0cm} p{1.0cm}p{0.005cm}
                p{1.0cm}p{1.0cm}p{1.0cm}p{1.0cm} p{1.0cm}p{0.005cm}
                p{1.0cm}p{1.0cm}p{1.0cm}p{1.0cm} p{1.0cm}
                }
\toprule
Dataset $\rightarrow$ & DailyDialog & &  &  &  &  & EmoryNLP &  &  & &  & & MELD & &  &  &  &  & IEMOCAP &  & &  &     \\ 
\midrule
Method $\downarrow$
&\normalsize Reward &\normalsize R@3 &\normalsize R@5 &\normalsize NDCG &\normalsize MRR &  &\normalsize Reward &\normalsize R@3 &\normalsize R@5 &\normalsize NDCG &\normalsize MRR &  &\normalsize Reward &\normalsize R@3 &\normalsize R@5 &\normalsize NDCG &\normalsize MRR &  &\normalsize Reward &\normalsize R@3 &\normalsize R@5 &\normalsize NDCG &\normalsize MRR \\ 
\midrule
0-shot & 0.37 & 0.47 & 0.56 & 0.84 & 0.48 & & 0.57 & 0.52 & 0.69 & 0.80 & 0.47 & & 0.70 & 0.53 & 0.69 & 0.81 & 0.49 & & 0.59 & 0.41 & 0.48 & 0.83 & 0.47 \\
ECoT   & 0.10 & 0.51 & 0.65 & 0.79 & 0.41 & & 0.57 & 0.45 & 0.68 & 0.77 & 0.39 & & 0.65 & 0.48 & 0.75 & 0.77 & 0.42 & & 0.51 & 0.28 & 0.41 & 0.78 & 0.33 \\  
PS     & 0.43 & 0.60 & 0.68 & 0.86 & 0.57 & & 0.61 & 0.56 & 0.73 & 0.81 & 0.47 & & 0.79 & 0.55 & 0.73 & 0.81 & 0.50 & & 0.63 & 0.45 & 0.52 & 0.83 & 0.48 \\
MP     & 0.40 & 0.56 & 0.63 & 0.85 & 0.53 & & 0.66 & 0.50 & 0.67 & 0.79 & 0.47 & & 0.78 & 0.51 & 0.66 & 0.80 & 0.47 & & 0.61 & 0.39 & 0.47 & 0.82 & 0.45 \\
\midrule
SFT    & \underline{0.55} & 0.79 & 0.86 & \underline{0.88} & \underline{0.70} & & 0.68 & \underline{0.59} & \underline{0.74} & \underline{0.83} & \textbf{0.51} & & 0.83 & 0.64 & 0.87 & \underline{0.84} & \underline{0.59} & & 0.68 & 0.51 & 0.56 & 0.84 & \underline{0.53} \\
EmoFSM    & 0.52 & 0.73 & 0.81 & 0.88 & 0.67 & & \underline{0.70} & 0.56 & 0.72 & 0.82 & \underline{0.51} & & \underline{0.84} & \underline{0.65} & \underline{0.88} & 0.83 & \textbf{0.59 }& & \underline{0.73} & \underline{0.52} & \underline{0.57} & \underline{0.84} & \textbf{0.54 }\\ 
\midrule
EMDP   & 0.33 & \textbf{0.83} & \underline{0.88} & 0.86 & 0.71 & & 0.44 & 0.46 & 0.63 & 0.75 & 0.47 & & 0.78 & 0.55 & 0.79 & 0.74 & 0.51 & & 0.53 & 0.26 & 0.44 & 0.69 & 0.36 \\
{\ModelName} 
       & \textbf{0.57} & \underline{0.82} & \textbf{0.92} & \textbf{0.92} & \textbf{0.72} & & \textbf{0.71} & \textbf{0.63} & \textbf{0.84} & \textbf{0.83} & 0.50 & & \textbf{0.86} & \textbf{0.69 }& \textbf{0.89} & \textbf{0.85} & 0.54 & & \textbf{0.81 }& \textbf{0.54} & \textbf{0.71} & \textbf{0.85} & 0.45 \\
\bottomrule
\end{tabular}}
\caption{Results on emotion determination.}
\label{tab:main_emo_result}
\end{table*}

\begin{table*}[htbp!]
\renewcommand{\arraystretch}{1.1}
\fontsize{14pt}{16pt}\selectfont
\centering
\resizebox{0.98\textwidth}{!}{
\begin{tabular}{p{2.4cm} 
                p{1.0cm}p{1.0cm}p{1.0cm}p{1.0cm}p{0.005cm}
                p{1.0cm}p{1.0cm}p{1.0cm} p{1.0cm}p{0.005cm}
                p{1.0cm}p{1.0cm}p{1.0cm} p{1.0cm}p{0.005cm}
                p{1.0cm}p{1.0cm}p{1.0cm} p{1.0cm}
                }
\toprule
Dataset $\rightarrow$ & DailyDialog &  &  &  &  & EmoryNLP &  &  &  & & MELD &  &  &  &  & IEMOCAP &  &  &     \\ 
\midrule
Method $\downarrow$
& B-2 & R-L & D-2 & CIDEr &  & B-2 & R-L & D-2 & CIDEr &  & B-2 & R-L & D-2 & CIDEr &  & B-2 & R-L & D-2 & CIDEr \\
\midrule
0-shot       & 3.53 & 11.48 & 40.81 & 3.63 &  & 2.08 & 8.85 & 60.62 & 2.52 & & 1.77 & 9.15 & 52.17 & 2.31 &  & 1.74 & 6.74 & 6.3 & 1.28 \\
ECoT         & 0.86 & 3.57 & 14.13 & 0.51 &  & 0.51 & 2.56 & 19.58 & 0.34 & & 0.47 & 2.41 & 15.65 & 0.63 &  & 0.5 & 2.47 & 9.14 & 0.71 \\ 
PS           & 2.40 & 6.73 & 35.97 & 1.41 &  & 1.56 & 5.58 & 53.69 & 1.10 &  & 1.53 & 5.15 & 44.58 & 0.98 &  & 1.16 & 3.59 & 3.51 & 0.30 \\
MP           & 1.30 & 4.97 & 13.4 & 1.17 &  & 0.99 & 4.33 & 33.57 & 0.79 &  & 0.98 & 4.36 & 28.56 & 0.75 &  & 0.76 & 3.17 & 6.93 & 0.21 \\
\midrule
SFT          & 6.27 & {21.81} & \underline{51.76} & 22.06 &  & 3.68 & 12.2 & \underline{12.35} & 11.93 & & 3.12 & \underline{10.8} & \textbf{21.09} & \underline{10.48} &  & 6.52 & 13.25 & 1.8 & \underline{36.83} \\
EmoFSM          & 6.32 & \underline{21.92} & 50.35 & 21.76 &  & \underline{3.89} & \underline{12.25} & 7.84 & \underline{12.26} & & \underline{3.85} & 10.55 & \underline{11.21} & \textbf{13.38} &  & 3.24 & 9.52 & \underline{38.86} & 13.15 \\ 
\midrule
EMDP         & \underline{6.66} & 20.25 & 51.08 & \underline{24.78} &  & 3.26 & 11.22 & 3.53 & 10.76 & & 2.61 & 9.89 & 7.66 & 8.27 &  & \underline{6.77} & \underline{16.35} & 18.7 & 35.27 \\
{\ModelName} & \textbf{9.11} & \textbf{25.34} & \textbf{54.72} & \textbf{41.73} &  & \textbf{4.39} & \textbf{12.89} & \textbf{16.82} & \textbf{14.23 }& & \textbf{3.89 }& \textbf{12.19} & 9.86 & 9.32 &  & \textbf{20.1 }& \textbf{31.65} & \textbf{42.04} & \textbf{38.89} \\
\bottomrule
\end{tabular}}
\caption{Results on Response Generation.}
\label{tab:main_response_result}
\end{table*}

\subsection{Baselines}

We compare to the following \text{prompting baselines}:
\noindent (1) 0-shot: directly inference the LLM generator, with the same context.\\
\noindent (2) ECoT \citep{li2024enhancingemotionalgenerationcapability}: uses the CoT prompt, which first generates the seeker's \textit{emotion}, then guides the generation of strategy and response. \\
\noindent (3) Plan-and-Solve (PS) \citep{wang-etal-2023-plan}: first prompts LLMs to generate a detailed plan outlining sub-goals and reasoning strategies, then executes the plan step-by-step to complete the solution.\\
\noindent (4) Metacognitive Prompting (MP) \citep{wang-zhao-2024-metacognitive}: it guides LLMs to perform structured self-reflection by generating, evaluating, and revising their own reasoning steps.


We also investigate these \textbf{supervised baselines}:

\noindent (5) SFT: Supervised fine-tuning on the emotion-annotated samples with the ECoT prompt.\\
\noindent (6) EmoFSM \citep{10.1007/978-981-92-1926-1_10}: The method guides the model by a finite state machine, with prompts based on inter-state transitions (context, emotion, strategy, response), then finetuning on the reformulated samples.\\
\noindent We finally include a \textbf{RL-based baseline}:\\
\noindent (7) EMDP \citep{sun2023dynamic}: Conduct a tabular Q-learning to determine the optimal policy on the emotional Markov decision process.

\subsection{Results}

\paragraph{Losses.} Figure~\ref{fig:training_curves} illustrates the training loss curves of {\ModelName}.
We observe that the training process is stable across all datasets, with the loss consistently decreasing and converging without oscillation, indicating the training stability preserved with the new types of loss and rewards.

\paragraph{Automatic metrics.} Tables \ref{tab:main_emo_result} and \ref{tab:main_response_result} report the results of the automatic metrics for emotion determination and response generation, respectively.
Across all four datasets, {\ModelName} consistently achieves the highest or near-highest performance on the reward and ranking metrics.
Compared with both zero-shot prompting methods and supervised baselines, our approach demonstrates a clear advantage in ranking emotionally appropriate actions higher, indicating more reliable emotional decision-making.
The results of Recall@3/5, NDCG, and MRR suggest that the learned Q-values capture meaningful relative preferences among candidate emotions rather than just optimizing for a single dominant action.

For response generation (Table \ref{tab:main_response_result}), {\ModelName} achieves better scores on BLEU-2, ROUGE-L, CIDEr, and Distinct-2.
This indicates that better emotional decision-making at the policy level translates into better responses.

\paragraph{Human evaluation.} We invited 10 interns as evaluators for the human evaluation. We sampled 50 dialogue sessions from each test set, and each evaluator independently scored all generated responses. The detailed annotation principles are shown in Appendix \ref{human eval}.

\begin{table*}[h!]
\vspace{-3.5mm}
\centering
\renewcommand{\arraystretch}{1.1}
\fontsize{9pt}{12pt}\selectfont
\resizebox{0.98\textwidth}{!}{%
\begin{tabular}{p{1.5cm}
    >{\centering\arraybackslash}p{1.2cm}>{\centering\arraybackslash}p{0.5cm}>{\centering\arraybackslash}p{1.2cm}>{\centering\arraybackslash}p{0.5cm}>{\centering\arraybackslash}p{1.2cm}>{\centering\arraybackslash}p{0.5cm}>{\centering\arraybackslash}p{1.2cm}
    >{\centering\arraybackslash}p{0.5cm}>{\centering\arraybackslash}p{1.2cm}>{\centering\arraybackslash}p{0.5cm}>{\centering\arraybackslash}p{1.2cm}>{\centering\arraybackslash}p{0.5cm}>{\centering\arraybackslash}p{1.2cm}>{\centering\arraybackslash}p{0.5cm}
    }
\toprule
& {\textbf{Fluency}} & \textbf{\;\, p} & \textbf{Emotion} & \textbf{\;\, p} & \textbf{Acceptance} & \textbf{\;\, p} & \textbf{Effectiveness} & \textbf{\;\, p} & \textbf{sensitivity} & \textbf{\;\, p} & \textbf{Alignment} & \textbf{\;\, p} & \textbf{Satisfaction} & \textbf{\;\, p} \\
\midrule
0-shot & 3.14(1.26) & <0.01 & 3.1(1.26) & <0.01 & 2.64(1.19) & <0.01 & 2.86(1.26) & <0.01 & 2.87(1.29) & <0.01 & 2.84(1.26) & <0.01 & 2.91(0.51) & <0.01 \\
ECoT & 3.12(1.28) & <0.01 & 3.12(1.32) & <0.01 & 2.73(1.19) & <0.01 & 2.63(1.15) & <0.01 & 3.05(1.23) & <0.01 & 2.72(1.21) & <0.01 & 2.89(0.45) & <0.01 \\
PS & 3.07(1.28) & <0.01 & 3.16(1.26) & <0.01 & 2.73(1.22) & <0.01 & 2.99(1.28) & <0.01 & 2.86(1.29) & <0.01 & 2.99(1.17) & <0.01 & 2.97(0.51) & <0.01 \\
MP & 3.17(1.21) & <0.01 & 3.14(1.26) & <0.01 & 2.71(1.26) & <0.01 & 2.63(1.23) & <0.01 & 2.74(1.31) & <0.01 & 2.92(1.22) & <0.01 & 2.89(0.52) & <0.01 \\

\midrule
SFT &  3.10(1.23) & <0.01 & 3.42(1.25) & <0.01 & 2.9(1.26) & <0.01 & 2.69(1.15) & <0.01 & 2.87(1.21) & <0.01 & 3.08(1.24) & <0.01 & 3.01(0.49) & <0.01 \\
EmoFSM & 3.37(1.27) & <0.01 & 3.43(1.23) & <0.01 & 2.99(1.29) & <0.01 & 3.05(1.31) & <0.01 & 3.05(1.21) & <0.01 & 3.19(1.29) & <0.05 & 3.18(0.53) & <0.01 \\
\midrule
EMDP & 3.43(1.19) & <0.01 & 3.36(1.30) & <0.01 & 2.81(1.22) & <0.01 & 2.65(1.23) & <0.01 & 3.02(1.2) & <0.01 & 3.12(1.28) & <0.01 & 3.07(0.47) & <0.01\\
\textbf{\ModelName} & \textbf{3.58}(1.30) &--& \textbf{3.74}(1.22) &--& \textbf{3.11}(1.31) &--& \textbf{3.01}(1.24) &--& \textbf{3.21}(1.30) &--& \textbf{3.22}(1.32) &--& \textbf{3.31}(0.48) &--\\
\bottomrule
\end{tabular}
}
\caption{Human evaluation of response quality on DailyDialog, EmoryNLP, MELD, and IMEOCAP.}
\label{tab:response_quaility}
\end{table*}

Averaged from the evaluators’ ratings, the final scores of each method are reported in Table \ref{tab:response_quaility}. Regarding statistical significance, we conducted t-tests on the average scores between different methods. The results of the significance tests are also included in Table \ref{tab:response_quaility}, under the null hypothesis $H_0: metric (\text{X}) > metric (\text{\ModelName})$. Results indicate that {\ModelName} achieves statistically significant improvements against most baselines. To indicate the consistence between different annotators, we include their correlation studies in Appendix \ref{inter-annotator agreement}.

\paragraph{Ablation study.} Table~\ref{tab:ablation_result} reports the results of our ablation studies. w/ SFT introduces an additional SFT stage before RL, which performs worse than directly applying RL. By inspecting the training curves, we find that such prior fine-tuning makes exploration more difficult, which in turn degrades the model’s capability.

Another ablation is to generate the Q-values by an extra MLP head (w/ head), which is trained from scratch. Although this method is adopted by many reward model implementations, in our study, it has degraded performances across all metrics, confirming that the Q-modeling adopted in {\ModelName} is more effective.


Removing the dialogue history (w/o history) or the emotion description (w/o desc) also results in performance drops. These results verify that both components contribute meaningfully to the emotional determination of our framework.

\begin{table*}[h!]
\renewcommand{\arraystretch}{1.1}
\fontsize{14pt}{16pt}\selectfont
\centering
\resizebox{0.98\textwidth}{!}{
\begin{tabular}{p{2.6cm} 
                p{0.9cm}p{0.9cm}p{0.9cm}p{0.9cm}p{0.9cm}p{0.1cm}
                p{0.9cm}p{0.9cm}p{0.9cm}p{0.9cm}p{0.9cm}p{0.1cm}
                p{0.9cm}p{0.9cm}p{0.9cm}p{0.9cm}p{0.9cm}p{0.1cm}
                p{0.9cm}p{0.9cm}p{0.9cm}p{0.9cm}p{0.9cm}}
\toprule
Dataset $\rightarrow$ & DailyDialog &  &  &  &  &  & EmoryNLP &  &  &  &  &  & MELD &  &  &  &  &  & IMEOCAP &  &  &  &  \\
\midrule
Method $\downarrow$
&\normalsize Reward &\normalsize R@3 &\normalsize R@5 &\normalsize NDCG &\normalsize MRR &  &\normalsize Reward &\normalsize R@3 &\normalsize R@5 &\normalsize NDCG &\normalsize MRR &  &\normalsize Reward &\normalsize R@3 &\normalsize R@5 &\normalsize NDCG &\normalsize MRR &  &\normalsize Reward &\normalsize R@3 &\normalsize R@5 &\normalsize NDCG &\normalsize MRR \\ 
 \midrule
\textbf{\ModelName} & \textbf{0.57} & 0.82 & \textbf{0.92} & \textbf{0.92} & \textbf{0.72} &  & \textbf{0.81}& 0.63 & \textbf{0.84} & 0.83 & \textbf{0.50} &  & \textbf{0.86} & \textbf{0.69} & \textbf{0.88} & \textbf{0.85} & \textbf{0.54} &  & \textbf{0.75} & \textbf{0.54} &\textbf{0.71} & \textbf{0.85} & \textbf{0.45} \\
\;w/ SFT & 0.45 & \textbf{0.83} & 0.92 & 0.92 & 0.72 &  & 0.69 & \textbf{0.66} & 0.83 & \textbf{0.85} & 0.54 &  & 0.81 & 0.65 & 0.90 & 0.84 & 0.49 &  & 0.73 & 0.52 & 0.68 & 0.85 & 0.43 \\
\;w/ head & 0.47 & 0.23 & 0.77 & 0.78 & 0.33 &  & 0.60 & 0.40 & 0.67 & 0.76 & 0.36 &  & 0.64 & 0.44 & 0.73 & 0.74 & 0.35 &  & 0.59 & 0.33 & 0.6 & 0.77 & 0.34\\
\midrule
\;w/o history  & 0.21 & 0.82 & 0.92 & 0.88 & 0.48 &  & 0.61 & 0.49 & 0.66 & 0.83 & 0.47 &  & 0.73 & 0.76 & 0.90 & 0.84 & 0.49 &  & 0.58 & 0.38 & 0.53 & 0.8 & 0.35 \\
\;w/o desc     & 0.33 & 0.82 & 0.92 & 0.92 & 0.72 &  & 0.67 & 0.60 & 0.82 & 0.83 & 0.50 &  & 0.75 & 0.70 & 0.90 & 0.85 & 0.51 &  & 0.46 & 0.49 & 0.66 & 0.84 & 0.41 \\ \bottomrule
\end{tabular}}
\caption{Results of ablation studies.}
\label{tab:ablation_result}
\end{table*}


\paragraph{Sensitivity analysis.}

\begin{figure}[!t]
\centering
  \includegraphics[width=0.55\linewidth]{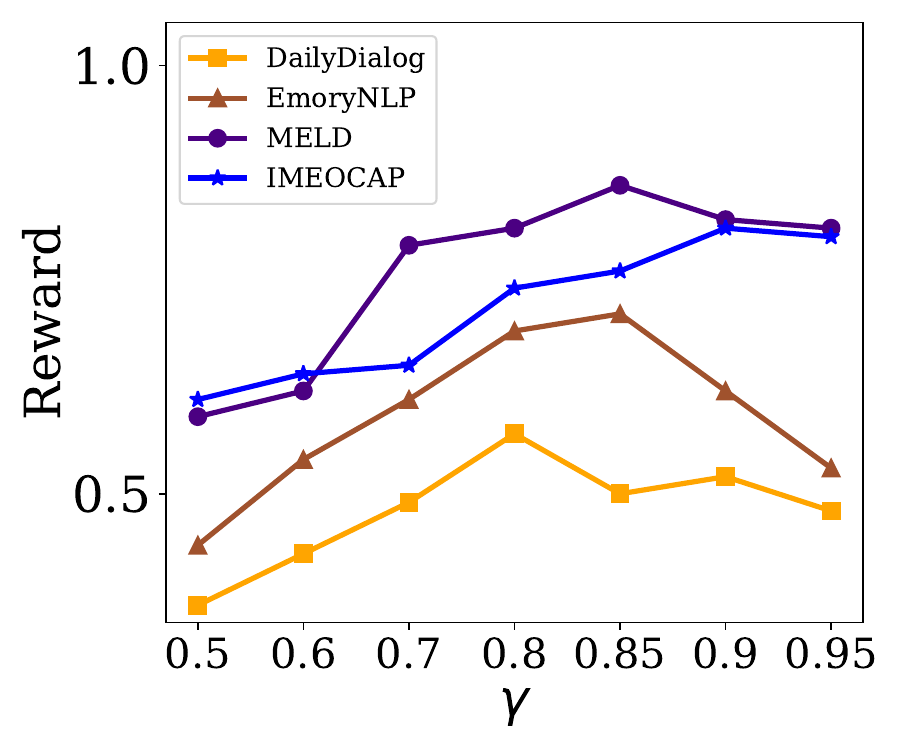} 
  \caption{Average rewards on different choices of $\gamma$.}
  \label{fig:reward_gamma}
  \vspace{-3mm}
\end{figure}

Figure \ref{fig:reward_gamma} analyzes the sensitivity of the average reward to the discount factor $\gamma$.
We observe that the optimal $\gamma$ varies across datasets and correlates with their average dialogue length. Datasets with longer conversational trajectories, such as IEMOCAP, favor larger $\gamma$ values, as long-term emotional consistency becomes more important. In contrast, DailyDialog, which consists of shorter interactions, reaches optimal performance at a smaller $\gamma$. Another sensitivity study on $w$ can be found in Appendix\ref{sensitivity}.

\subsection{Downstream TTS experiment}

To further evaluate the practical utility of our emotional decision model, we conduct a downstream emotional TTS experiment using CosyVoice2 \citep{du2024cosyvoice}. For the inference results of all methods, we consistently employ the "instruct inference" method from CosyVoice2 for emotional speech synthesis. Besides the baselines previously introduced, we also include a prompting baseline called CoCT \citep{10.1007/978-981-95-7078-2_23}, which encourages the LLM to first propose a concept (\textit{e.g.}, the emotion), then decodes the detailed response.

We evaluate the quality of the synthesized speech by SpeechBERTScore (BERT) \citep{saeki2024speechbertscore} and PESQ \citep{recommendation2001perceptual}. SpeechBERTScore computes token-level similarity between the generated speech and the reference utterance in a shared embedding space, while PESQ is a reference-aware objective metric to evaluate the perceptual speech quality. It assumes the generated and reference speech signals are time-aligned. 

As shown in Figure \ref{fig:paradigm} Setting B, ERC can not produce emotion until the entire response is generated. Therefore, this emotion label can not be utilized in the streaming TTS, which requires the emotion provided before the speech synthesis. In contrast, our approach determines emotion prior to the response generation, as indicated by Figure \ref{fig:paradigm} Setting C. While Setting C supports streaming output, it may entail a compromise in emotional accuracy. To ensure that our method can maintain high output quality even without the steaming setting, we compare the results of various methods on both Setting B and Setting C. As demonstrated in Table \ref{tab:tts_metrics}, {\ModelName} outperforms other methods in terms of generation quality under Setting B, whereas the other methods experience a significant decline under Setting C. Audio demonstrations are available on our project website.




\subsection{Discussion}


\paragraph{Emotion transition matrix.}

Figure~\ref{fig:training_curves} visualizes the state-action transition matrices, in which grid (i,j) indicates the current emotion $i$ is followed by a transition to emotion $j$. Most transitions happen on the diagonal grids and their adjacent grids, which aligns well with the topology of Figure \ref{Plutchik wheel}. In contrast, transitions between opposite emotions are consistently suppressed, indicating that the learned policy internalizes the theoretical constraints encoded by the Plutchik score.

\begin{figure}[htbp!]
    \centering
    \includegraphics[width=0.42\linewidth]{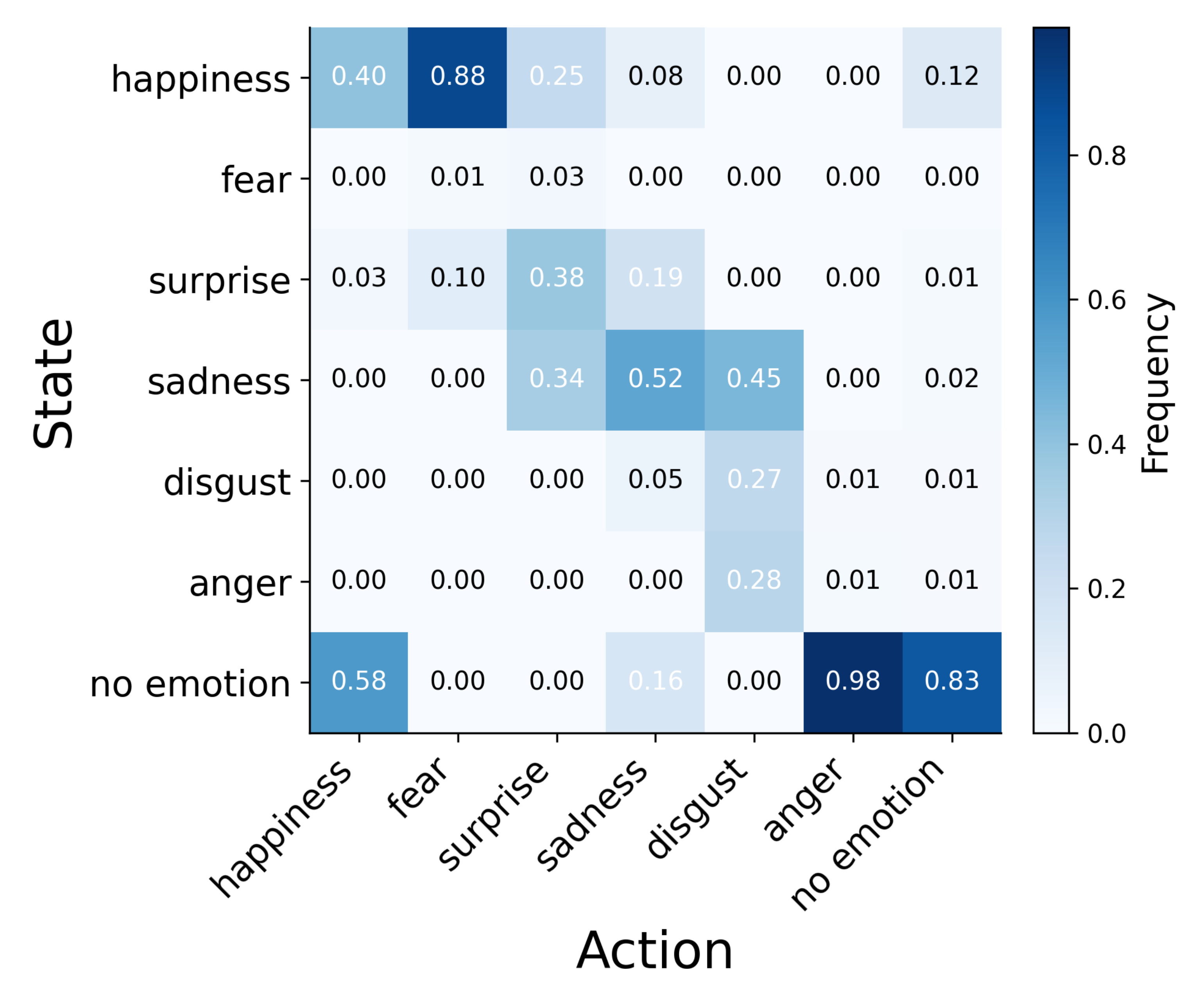}
    \hspace{0.05in}
    \includegraphics[width=0.42\linewidth]{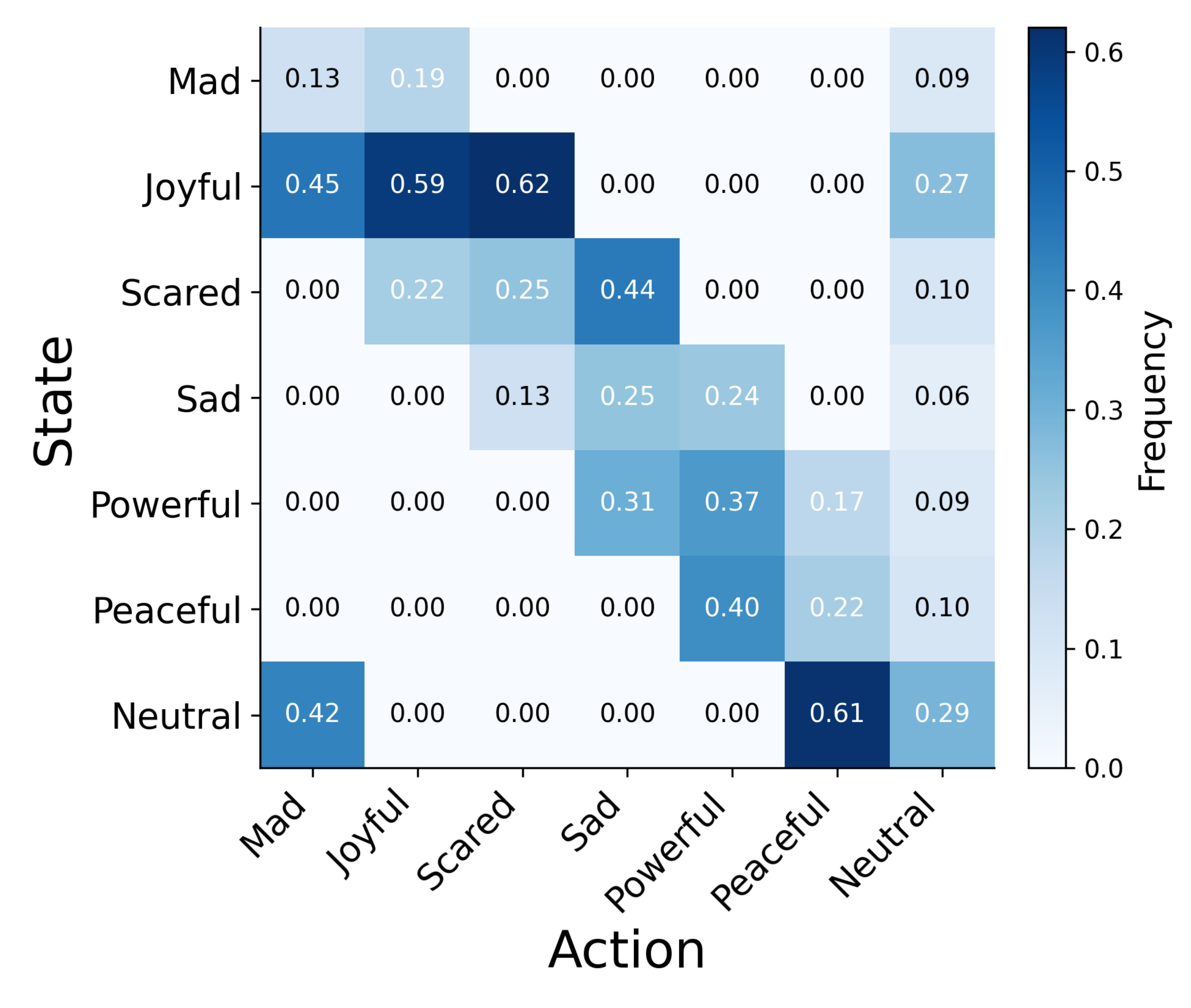}
    \vspace{-3.5mm}
    \includegraphics[width=0.42\linewidth]{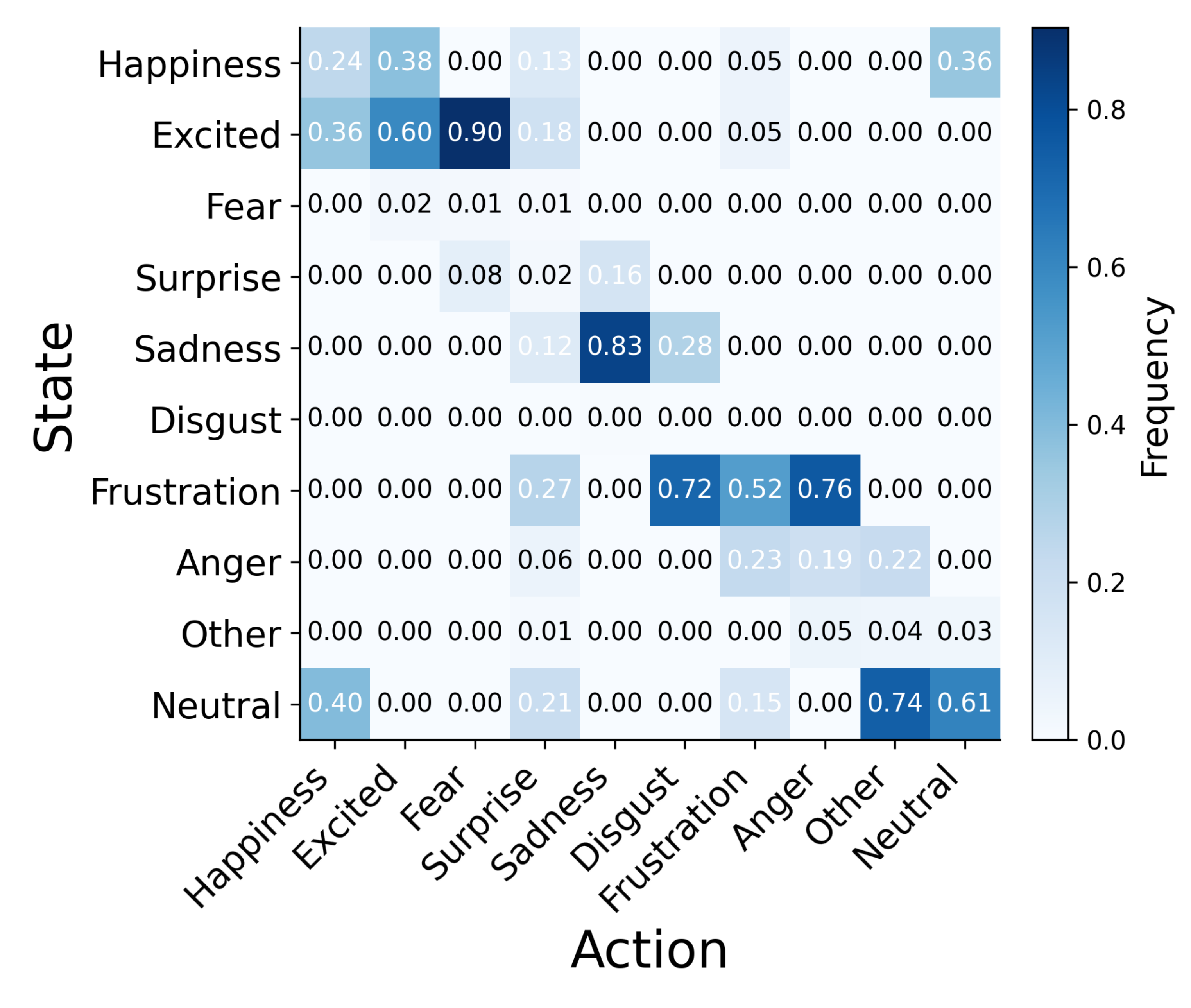}
    \hspace{0.05in}
    \includegraphics[width=0.42\linewidth]{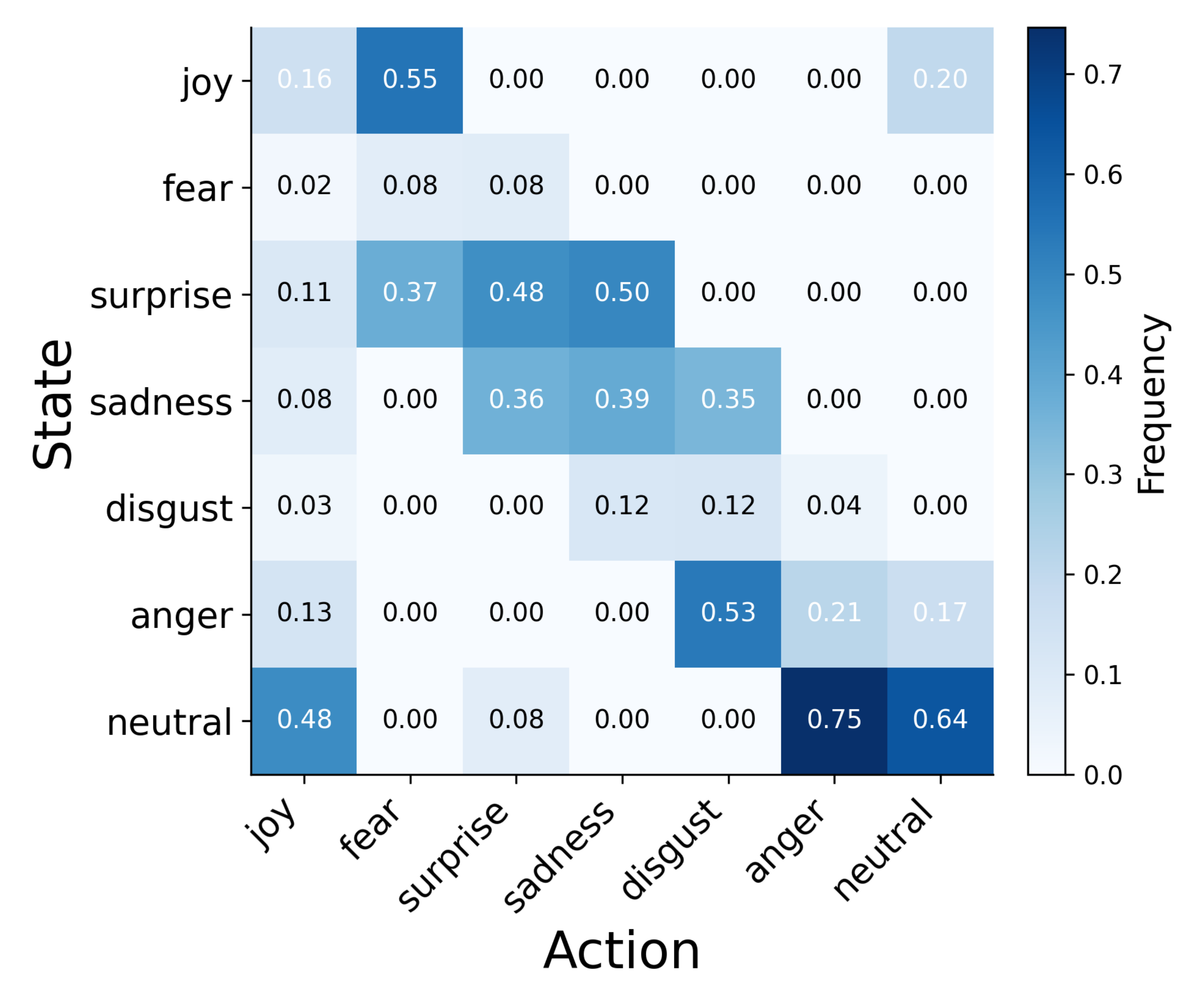}
    \caption{Emotional transition matrices of {\ModelName}. Upper-Left: DailyDialog; Upper-Right: EmoryNLP; Bottom-Left: IMEOCAP; Bottom-Right: MELD.}
    \label{fig:training_curves}
    \vspace{-3.5mm}
\end{figure}

\paragraph{Empirical verification of methodology.} Guided by the Plutchik-driven rewards, we further validate the Q-function, which drives the conversation agent, exhibits similar patterns as depicted in the Plutchik theory. We validate this consistency by investigating the occurrence of emotion-behavior transitions, in which the original theory depicts the typical pattens, as detailed in Appendix \ref{sec:emotion-behavior-transition}. Accordingly, we calculate the \textbf{the ratio of expected transitions} by 
\begin{equation*}
    R = \frac{\text{\# typical emotion-behavior transitions}}{\text{\# all emotion-behavior transitions}}
\end{equation*}

\begin{figure}[htbp!]
    \centering
    \includegraphics[width=0.6\linewidth]{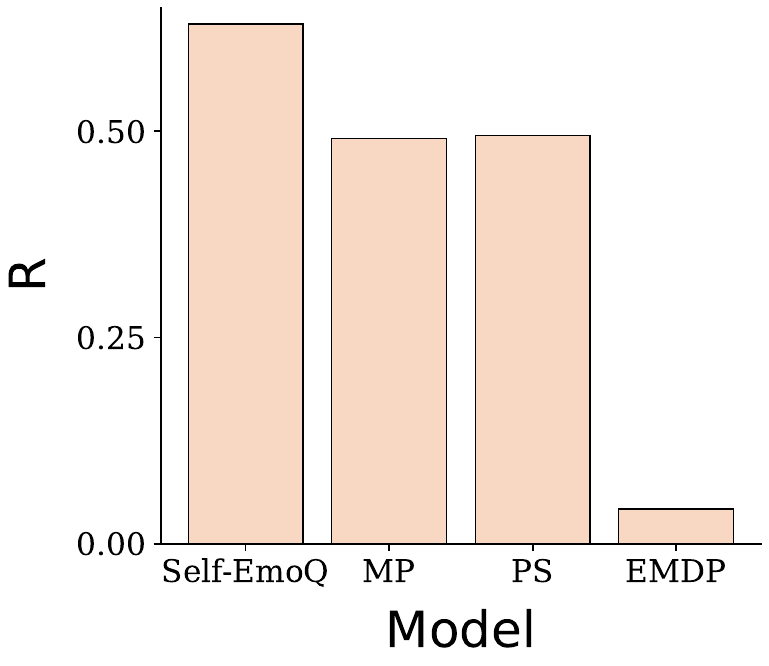}
    \vspace{-3.5mm}
    \caption{Ratio of expected transitions on MELD.}
    \label{fig:plutchik}
    \vspace{-3.5mm}
\end{figure}

In Figure \ref{fig:plutchik}, our model exhibits higher $R$ than baselines, indicating that our agent, driven by Plutchik's rewards, finally aligns well to the behavioral patterns depicted by the Plutchik theory.

To validate the GPT-based Plutchik scoring, we also conduct human annotations on a subset of these rewards, showing a high level correlation. Detailed analysis is in Appendix \ref{Evaluation of GPT Scoring}.

\begin{table*}[ht!]
\centering
\small
\begin{tabular}{lccccccccc}
\toprule
\multirow{2}{*}{\textbf{Method}} & \multirow{2}{*}{\textbf{Setting}} & \multicolumn{2}{c}{\textbf{DailyDialog}} & \multicolumn{2}{c}{\textbf{EmoryNLP}} & \multicolumn{2}{c}{\textbf{MELD}} & \multicolumn{2}{c}{\textbf{IEMOCAP}} \\
\cmidrule(l){3-4} \cmidrule(l){5-6} \cmidrule(l){7-8} \cmidrule(l){9-10}
 &  & BERT & PESQ & BERT & PESQ & BERT & PESQ & BERT & PESQ \\
 \midrule
\multirow{2}{*}{0-shot} & B & 0.46 & 1.17 & 0.46 & 1.21 & 0.46 & 1.19 & 0.44 & 1.21 \\
 & C & 0.33 & 1.01 & 0.37 & 1.22 & 0.29 & 0.97 & 0.31 & 1.07 \\
\multirow{2}{*}{CoCT} & B & 0.48 & 1.17 & 0.46 & 1.24 & 0.47 & 1.22 & 0.46 & 1.21 \\
 & C & 0.34 & 1.03 & 0.36 & 1.23 & 0.31 & 1.01 & 0.32 & 1.09 \\
\multirow{2}{*}{ECoT} & B & 0.43 & 1.26 & 0.41 & 1.29 & 0.41 & 1.21 & 0.42 & 1.26 \\
 & C & 0.32 & 1.09 & 0.32 & 1.27 & 0.30 & 1.03 & 0.30 & 1.14 \\
\multirow{2}{*}{MP} & B & 0.42 & 1.25 & 0.40 & 1.31 & 0.41 & 1.27 & 0.42 & 1.26 \\
 & C & 0.29 & 1.07 & 0.29 & 1.31 & 0.27 & 1.06 & 0.19 & 1.13 \\
\multirow{2}{*}{PS} & B & 0.45 & 1.16 & 0.44 & 1.20 & 0.44 & 1.18 & 0.44 & 1.22 \\
 & C & 0.40 & 1.09 & 0.31 & 1.22 & 0.31 & 0.97 & 0.31 & 1.09 \\
  \midrule
\multirow{2}{*}{SFT} & B & 0.50 & 1.26 & 0.48 & 1.23 & 0.51 & 1.22 & 0.47 & 1.21 \\
 & C & 0.39 & 1.07 & 0.36 & 1.26 & 0.38 & 0.98 & 0.35 & 1.11 \\
\multirow{2}{*}{EmoFSM} & B & 0.53 & 1.27 & 0.50 & 1.26 & \textbf{0.55} & 1.28 & 0.49 & 1.24 \\
 & C & 0.41 & 1.06 & 0.38 & 1.24 & 0.40 & 1.09 & 0.38 & 1.12 \\
  \midrule
\multirow{2}{*}{EMDP} & B & 0.53 & 1.25 & 0.48 & 1.26 & 0.51 & 1.20 & 0.48 & 1.27 \\
 & C & 0.39 & 1.04 & 0.40 & 1.23 & 0.41 & 1.01 & 0.36 & 1.13 \\
\multirow{2}{*}{\textbf{\ModelName}} & B & \textbf{0.54} & \textbf{1.28} & \textbf{0.50} & \textbf{1.28} & 0.53 & \textbf{1.29} & \textbf{0.52} & \textbf{1.28} \\
 & C & \textbf{0.54} & \textbf{1.28} & \textbf{0.50} & \textbf{1.28} & \textbf{0.53} & \textbf{1.29} & \textbf{0.52} & \textbf{1.28}\\
 \bottomrule
\end{tabular}
\caption{Results of downstream speech synthesis. `Setting' refers to the paradigms depicted by Figure \ref{fig:paradigm} (B) or (C).}
\label{tab:tts_metrics}
\end{table*}

\paragraph{Good case.} We provide a typical case on IMEOCAP in Table~\ref{tab:good_cases}. More cases can be found in Appendix\ref{more case}. The speech waves are shown on the website.

\section{Related Work}

\subsection{Emotion Cognition}

Emotion Recognition in Conversation (ERC) has been widely studied. \citet{poria2017context} improved ERC accuracy by utilizing contextual information.
DialogueRNN \citep{majumder2019dialoguernn} and DialogueGCN \citep{ghosal2019dialoguegcn} further improved performance by modeling speaker states and inter-speaker dependencies. Building upon LLMs, InstructERC reframes ERC as a generative task and integrates multi-granularity supervision to reach state-of-the-art results \citep{lei2023instructerc}. \citet{shen2025coe} further incorporates ERC with auxiliary reasoning and speaker-aware tasks. However, ERC can only provide the emotion cognition after the text generation, which can not be directly applied on real-time streaming conversation pipelines.

\subsection{Emotion Prediction}

Emotion Prediction in Conversation (EPC) forecasting the interlocutor’s future emotional state before the next utterance emerges. Early researches leverage neural approaches \citep{bothe2017dialogue}, probabilistic models \citep{sun2019dynamic}, and multimodal acoustic–text fusion \citep{li2018inferring}. Later studies improve the prediction by modeling emotional dynamics with variational methods \citep{zhang2021predicting}, incorporating uncertainty-aware \citep{han2021exploring}, or structured temporal decoding \citep{yeh2020dialogical}. Recently, contrastive learning, pseudo-utterance generation, and multi-modal feature fusion further improve representation quality and prediction robustness \citep{xie2025pseudo,shi2024emotional,ju2023real}. Nevertheless, these EPC studies share a critical limitation: they are data-driven methods which means their emotion forecasts are limited by dataset annotations. In contrast, our study conducts strategic emotional decision-making, filling the gap where EPC methods fail to explore unusual emotions, and their ignorance on long-term optimization.


\begin{table}[htbp!]
\vspace{-2.5mm}
    \centering
    \small
    \resizebox{\columnwidth}{!}{%
    \renewcommand{\arraystretch}{1.5}
\begin{tabular}{c| l}
        \Xhline{2\arrayrulewidth}
        \multicolumn{2}{c}{\textbf{Case of IMEOCAP}}  \\ 

        \hline
        History & \makecell[l]{...\\male: (\textit{Sadness}) I guess that included you.\\
female: (\textit{Neutral}) And do you still feel that way?\\
male: (\textit{Neutral}) I want you now, Annie.\\
female: (\textit{Anger}) Because you mustn't feel\\ that way anymore because you have a right\\ to whatever you have, Chris, do you \\understand, everything\\
female: (\textit{Anger}) that includes me.  And your\\ money there's nothing wrong in your money.  \\Your father put hundreds of planes in the air.}  \\
        \hline
        0-shot & \makecell[l]{(\textit{Happiness}) Oh, Chris, I've missed you too.}    \\
        \hline
        EmoFSM & \makecell[l]{(\textit{Neutral}) Yes, thank you.}     \\ 
        \hline
        \textbf{\ModelName} & \makecell[l]{(\textit{Anger}) You can't think like that.}     \\ 
        \hline
        \Xhline{2\arrayrulewidth}
        
    \end{tabular}
    }
    \caption{Good Cases of {\ModelName}.} 
    \label{tab:good_cases}
    \vspace{-3.5mm}
\end{table}

\subsection{Emotional Text-to-Speech}


Most emotional TTS methods have a zero-shot paradigm, such as ELaTE \citep{kanda2024making} and EmoCtrl-TTS \citep{wu2024laugh}, improve controllability of laughter, prosody, or arousal–valence values, yet are reactive frameworks. Word-level or vector-based approaches \citep{wang2025word,liang2025ece,cho2025emosphere++,gao2025emo} enhance granularity and intensity control but treat emotion as static input. RL-based or prompt-guided methods \citep{yang2025rlaif,gao2025prompt} also focus on expressiveness and diversity without modeling emotion as a temporally planned policy, and large-scale systems like CosyVoice 2/3 \citep{du2024cosyvoice,du2025cosyvoice} follow a similar zero-shot paradigm.
Overall, the field lacks mechanisms that treat emotion as a planned trajectory. Our method directly addresses this gap by introducing an explicit emotional planning module.

\section{Conclusion}

We propose {\ModelName}, an emotion-planning dialogue framework that determines its self-emotion before response generation, readily driving LLM and Emo-TTS in the streaming paradigm. Based on Plutchik's Wheel of Emotion, we introduce an extra theory-driven reward that allows the agent not only to imitate the dataset pattern, but also to align with the human emotional behaviors directly. We solve the problem by conducting the value-based RL on the LLM-based module, with Q-value output by averaging output token logprobs. The planner optimizes the long-term return, as the play-and-plug module before the LLM and TTS. Experiments show that our method outperforms other baselines on multiple datasets, with reasonable state-action and emotion-behavior transitions.


\clearpage
\newpage

\section*{Limitations}

Our framework relies on large language models for reward scoring, which introduces additional computational overhead and potential bias in emotion evaluation. Moreover, emotion planning is currently restricted to a discrete set of emotions defined by Plutchik’s theory, which may limit the representation of more fine-grained or mixed emotional states in real-world interactions.

\bibliography{main}

\appendix

\section{Further Implementation Details}

\subsection{Prompt of Plutchik Scoring}
\label{Plutchik score}

Based on the theory of \textbf{Plutchik's Wheel of Emotion}, we evaluate the \textbf{Plutchik Score} $r_{\text{Plu}}$ by GPT-4o , using the evaluation prompt below.

\tcbset{
  colframe=black!75!white,
  colback=gray!5!white,
  boxrule=0.5pt,
  arc=2mm,
  left=1mm, right=1mm, top=1mm, bottom=1mm,
  fonttitle=\bfseries,
  before skip=5pt, after skip=5pt
}
\begin{tcolorbox}[breakable,title=Plutchik Score Prompt]
\textbf{You are an emotion evaluation module grounded in Plutchik’s Wheel of Emotion.
}
\textbf{Plutchik’s theory defines eight emotions:}\\
$\{Joy, Trust, Fear, Surprise, Sadness,\\ Disgust, Anger, Anticipation\}$.

\textbf{Note that these eight emotions are organized in a specific order and exhibit opposite and adjacent relationships. Specifically, $Joy$ and $Anticipation$ are also adjacent.}\\
\textbf{There are four pairs of opposite relationships:
}\\
$(Joy,Sadness)$\\
$(Trust,Disgust)$\\
$(Fear, Anger)$\\
$(Surprise, Anticipation)$\\

\textbf{According to Plutchik’s theory:}

\textbf{1. Each emotion is associated with typical behaviors and functions:
}\\
    - $Joy$: Courting, mating; Reproduction\\
   - $Trust$: Grooming, sharing; Affiliation\\
   - $Fear$: Running, or flying away;Protection\\
   - $Surprise$: Stopping, alerting; Orientation\\
   - $Sadness$: Crying for help; Reintegration\\
   - $Disgust$: Vomiting, pushing away; Rejection\\
   - $Anger$: Biting, hitting; Destruction\\
   - $Anticipation$: Examining, mapping; Exploration\\

\textbf{2. Emotional transitions follow structured relationships:}\\
   - Transitions between \textbf{adjacent} emotions are generally more natural.\\
   - Transitions between \textbf{opposite} emotions are less plausible unless mediated.\\
   - Emotional responses should not abruptly contradict the user's emotional state.\\

Your task is to evaluate the system response according to these principles.

\textbf{History:} \{$h$\} \\   
\textbf{\  User's emotion:} \{$e_u$\} \textbf{\  Query:} \{$query$\} \\
\textbf{\  System's emotion:} \{$e_s$\} \textbf{\  Response:} \{$response$\} \\

\textbf{Evaluate the system response using the following criteria. For each criterion, assign an integer score from 0 to 5 (0 = completely inappropriate, 5 = highly appropriate).
}\\

\textbf{1. Emotion Alignment:}\\
To what extent does the response clearly express the target emotional state?

\textbf{2. Emotion Transition Plausibility:}\\ Is the emotional transition from the user's emotion to the system's target emotion reasonable according to Plutchik’s emotional structure?

\textbf{3. Emotion–Function Consistency:}\\ Does the response exhibit the typical behavioral function associated with
   the target emotion?
   
\textbf{Your response needs to follow the following format:}\\
$\{
  alignment: int,\\
  transition: int,\\
  function: int,
\}$
\end{tcolorbox}

\subsection{Typical Emotion-Behavior Transitions}
\label{sec:emotion-behavior-transition}

The theory of \textit{Plutchik’s Wheel of Emotions} \citep{plutchik1982psychoevolutionary} identifies eight primary emotions: \textbf {joy}, \textbf{trust}, \textbf{fear}, \textbf{sadness}, \textbf{disgust}, \textbf{anger}, \textbf{anticipation}, and \textbf{surprise}; alongside eight secondary emotions, which are derived from combinations of primary ones. \textit{Plutchik} (we use it to abbreviate the theory for the rest of paper) also defines reasonable conversions from specific emotions to behaviors (and responding functions, which provides a higher-level abstraction for behaviors), as detailed shown in Table \ref{tab:Plutchik_conversions}.

\begin{table}[t!]
\centering
\small
\begin{tabular}{lll}
\toprule
   Emotion     &  Behavior   &  Function    \\ 
\midrule
Fear, Terror & Withdrawing & Protection \\ 
Anger, Rage & Attacking; Biting & Destruction \\
Joy, Ecstasy & Mating; Possessing & Reproduction \\
Sadness, Grief & Crying for Help & Reintegration \\
Acceptance & Pair Bonding & Incorporation \\ 
Disgust & Vomiting; Defecating & Rejection \\ 
Expectancy & Examining; Mapping & Exploration \\ 
Surprise & Stopping; Freezing & Orientation \\ 
\bottomrule
\end{tabular}
\caption{Typical transitions from emotional states to behaviors, as specified in Plutchik's Wheel of Emotion \citep{plutchik1982psychoevolutionary}.}
\label{tab:Plutchik_conversions}
\end{table}

\subsection{Metric Details on Emotion Determination}
\label{detail of metrics1}


\paragraph{Recall.} Given a dialogue state (or query) $s$, let $\mathcal{R}_K$ denote the top-$K$ ranked predictions produced by the model, and let $\mathcal{G}$ denote the set of relevant (ground-truth) emotion labels. The Recall@K is defined as:
\begin{equation}
\text{Recall@}K = \frac{|\mathcal{R}_K \cap \mathcal{G}|}{|\mathcal{G}|}.
\end{equation}

\paragraph{MRR.} The Mean Reciprocal Rank (MRR) \citep{voorhees-tice-2000-trec} evaluates the rank position of the first relevant prediction. Let $r$ denote the rank of the first relevant item in the predicted list (and $r=\infty$ if no relevant item is retrieved). The reciprocal rank is defined as:
\begin{equation}
\text{RR} = \frac{1}{r},
\end{equation}
and MRR is computed as the average over all $N$ dialogue states:
\begin{equation}
\text{MRR} = \frac{1}{N} \sum_{i=1}^{N} \frac{1}{r_i}.
\end{equation}

\paragraph{NDCG.} The Normalized Discounted Cumulative Gain at cutoff $p$ ($\mathrm{NDCG}_p$) \citep{jarvelin2002cumulated} accounts for graded relevance and ranking positions. Given the relevance score $\mathrm{rel}_i$ of the item ranked at position $i$, the discounted cumulative gain is defined as:
\begin{equation}
\mathrm{DCG}_p = \sum_{i=1}^{p} \frac{2^{\mathrm{rel}_i} - 1}{\log_2(i+1)}.
\end{equation}
The normalized DCG is obtained by:
\begin{equation}
\mathrm{NDCG}_p = \frac{\mathrm{DCG}_p}{\mathrm{IDCG}_p},
\end{equation}
where $\mathrm{IDCG}_p$ denotes the DCG at cutoff $p$ under the ideal ranking.

\subsection{Metric Details on Response Generation}
\label{detail of metrics2}


\paragraph{BLEU-2.} B-2\citep{papineni2002bleu} first compute the geometric average of the modified $n$-gram precisions, $p_n$, using $n$-grams up to length $N$ and positive weights $w_n$ summing to one.

Next, let $c$ be the length of the prediction and $r$ be the reference length. The BP and BLEU-2 are computed as follows.

\begin{equation}
    \mathrm{BP}=\left\{\begin{array}{ll}
1 & \text { if } c>r \\
e^{(1-r / c)} & \text { if } c \leq r
\end{array} .\right.
\end{equation}

\begin{equation}
    \mathrm{BLEU}=\mathrm{BP} \cdot \exp \left(\sum_{n=1}^N w_n \log p_n\right) .
\end{equation}

\paragraph{Rouge-L.} R-L\citep{lin2004rouge} propose using LCS-based F-measure to estimate the similarity between two summaries $X$ of length $m$ and $Y$ of length $n$, assuming $X$ is a reference summary sentence and $Y$ is a candidate summary sentence, as follows:

\begin{equation}
\begin{aligned}
& R_{l c s}=\frac{L C S(X, Y)}{m} \\
& P_{l c s}=\frac{L C S(X, Y)}{n} \\
& F_{l c s}=\frac{\left(1+\beta^2\right) R_{l c s} P_{l c s}}{R_{l c s}+\beta^2 P_{l c s}}
\end{aligned}
\label{rouge_l}
\end{equation}

Where $\operatorname{LCS}(X, Y)$ is the length of a longest common subsequence of $X$ and $Y$, and $\beta=P_{l c s} / R_{\text {lcs }}$ when $\partial F_{l c s} / \partial R_{l c s}=\partial F_{l c s} / \partial P_{l c s}$. In DUC, $\beta$ is set to a very big number $(\rightarrow \infty)$. Therefore, the LCS-based F-measure, \textit{i.e.}, Equation \ref{rouge_l}, is Rouge-L. 

\paragraph{CIDEr.} The CIDEr$_n$ \citep{vedantam2015cider} score for $n$-grams of length $n$ is computed using the average cosine similarity between the candidate sentence and the reference sentences, which accounts for both precision and recall:
\begin{equation}\label{eq:2}
CIDEr_n(c_i, S_i) = \frac{1}{m}\sum_j \frac{\textbf{g}^\textbf{n}(c_{i})\cdot \textbf{g}^\textbf{n}(s_{ij})}{\|\textbf{g}^\textbf{n}(c_{i})\|\|\textbf{g}^\textbf{n}(s_{ij})\|},
\end{equation}
where $\textbf{g}^\textbf{n}(c_{i})$ is a vector formed by $g_k(c_{i})$ corresponding to all $n$-grams\ of length $n$ and $\|\textbf{g}^\textbf{n}(c_{i})\|$ is the magnitude of the vector $\textbf{g}^\textbf{n}(c_{i})$. Similarly for $\textbf{g}^\textbf{n}(s_{ij})$.

Higher order (longer) $n$-grams are used to capture grammatical properties as well as richer semantics. \citep{vedantam2015cider} combine the scores from $n$-grams of varying lengths as follows:

\begin{equation} \label{eq:3}
CIDEr(c_i, S_i) = \sum_{n=1}^N w_n CIDEr_n(c_i, S_i),
\end{equation}
Empirically, Vedantam et al.\citep{vedantam2015cider} found that uniform weights $w_n=1/N$ work the best. So, we also use $N$ = 4.

\paragraph{Dist-2.} \citet{li2015diversity} report the degree of diversity by calculating the number of distinct unigrams and bigrams in generated responses.
The value is scaled by the total number of generated tokens to avoid favoring long sentences:
\begin{equation} \label{eq:4}
Dist(n) = \frac{Count(unique\ n-gram)}{Count(n-gram)}
\end{equation}

\subsection{Dataset Details}\label{introduction dataset}

\textbf{DailyDialog} consists of dyadic conversations covering various topics of daily-life and is designed to reflect natural human communication. Each utterance is annotated with both emotion categories and dialogue acts. The emotion labels include anger, disgust, fear, happiness, sadness, surprise, and neutral. 

\textbf{MELD} is an extension of the EmotionLines dataset and is a multimodal corpus
collected from the TV series \emph{Friends}. It contains over 1,400 dialogues and 13,000 utterances, where each utterance is annotated with both emotion and sentiment labels. 

\textbf{EmoryNLP} is also derived from the TV series \emph{Friends}, but differs from MELD in the selection of scenes and the definition of emotion categories. It consists of multi-party conversations with utterances annotated into seven emotion classes.

\textbf{IEMOCAP} is a widely used benchmark dataset in affective computing, consisting of approximately 12 hours of audiovisual recordings collected from dyadic interactions. Each conversation is segmented into utterances annotated with both categorical emotion labels (e.g., anger, happiness, sadness, and neutral) and continuous Valence-Arousal values.

\subsection{Principle of Human Scoring}
\label{human eval}

We start with the criteria proposed by \citet{kang-etal-2024-large}. The human evaluation is aimed to align with the ultimate purpose of emotional dialogue, the seeker's \textit{satisfaction}. To achieve this, the supporter's behavior can be further classified into the following criteria:

\noindent \textit{Acceptance}: Does the seeker accept without discomfort;

\noindent \textit{Effectiveness}: Is it helpful in shifting negative emotions or attitudes towards a positive direction; 

\noindent \textit{Sensitivity}: Does it take into consideration the general state of the seeker. Furthermore, to clarify the capability of LLMs to align strategy and responses, we include Alignment.

To achieve a more elaborate assessment, we consider three more dimensions addressing the generation quality:

\noindent \textit{Fluency}: the level of fluency of response.

\noindent \textit{Emotion}: the emotional intensity of response which could affect the seeker's emotional state.

\noindent \textit{Interesting}: Whether the response can arouse the seeker's interest and curiosity, presenting unique ideas, vivid expressions or engaging elements that capture the seeker's attention and make the interaction more appealing.

We engage our interns as human evaluators to rate the models according to these multiple aspects, namely Fluency, Emotion, Interesting, and Satisfaction, with Satisfaction covering Acceptance, Effective, Sensitivity, and Satisfaction itself. \\
Throughout this evaluation process, we strictly comply with international regulations and ethical norms, ensuring that all practices conform to the necessary guidelines regarding participant involvement and data integrity.\\
Evaluators are required to independently evaluate each sample in strict accordance with the pre - established criteria. By adhering to these principles, the evaluation process maintains objectivity, standardization, and consistency, thus enhancing the overall quality and credibility of the evaluation results. \\

\begin{table*}[t!]
\centering
\small
\begin{tabular}{l|lllllllllll}
\toprule
 & \textbf{eval\_0} & \textbf{eval\_1} & \textbf{eval\_2} & \textbf{eval\_3} & \textbf{eval\_4} & \textbf{eval\_5} & \textbf{eval\_6} & \textbf{eval\_7} & \textbf{eval\_8} & \textbf{eval\_9} & \textbf{eval\_9} \\
 \midrule
\textbf{eval\_0} & 1.00 & 0.83 & 0.81 & 0.76 & 0.82 & 0.81 & 0.80 & 0.81 & 0.81 & 0.82 & 0.82 \\
\textbf{eval\_1} & 0.83 & 1.00 & 0.83 & 0.78 & 0.85 & 0.83 & 0.83 & 0.83 & 0.84 & 0.84 & 0.84 \\
\textbf{eval\_2} & 0.81 & 0.83 & 1.00 & 0.78 & 0.83 & 0.82 & 0.82 & 0.82 & 0.83 & 0.82 & 0.82 \\
\textbf{eval\_3} & 0.76 & 0.78 & 0.78 & 1.00 & 0.78 & 0.77 & 0.76 & 0.77 & 0.78 & 0.78 & 0.78 \\
\textbf{eval\_4} & 0.82 & 0.85 & 0.83 & 0.78 & 1.00 & 0.83 & 0.83 & 0.83 & 0.83 & 0.82 & 0.82 \\
\textbf{eval\_5} & 0.81 & 0.83 & 0.82 & 0.77 & 0.83 & 1.00 & 0.81 & 0.83 & 0.83 & 0.82 & 0.82 \\
\textbf{eval\_6} & 0.80 & 0.83 & 0.82 & 0.76 & 0.83 & 0.81 & 1.00 & 0.81 & 0.81 & 0.81 & 0.81 \\
\textbf{eval\_7} & 0.81 & 0.83 & 0.82 & 0.77 & 0.83 & 0.83 & 0.81 & 1.00 & 0.82 & 0.83 & 0.83 \\
\textbf{eval\_8} & 0.81 & 0.84 & 0.83 & 0.78 & 0.83 & 0.83 & 0.81 & 0.82 & 1.00 & 0.82 & 0.82 \\
\textbf{eval\_9} & 0.82 & 0.84 & 0.82 & 0.78 & 0.82 & 0.82 & 0.81 & 0.83 & 0.82 & 1.00 & 1.00\\
\bottomrule
\end{tabular}
\caption{The Cohen’s Kappa Matrix among Evaluators of Fluency Score.}
\label{tab:Kappa_matrix}
\end{table*}

The detailed manual scoring criteria are as follows:
\begin{itemize}
\item Fluency:

1: The sentence is highly incoherent, making it extremely difficult to understand and failing to convey a meaningful idea.

2: The sentence has significant incoherence issues, with only parts of it making sense and struggling to form a complete thought.

3: The sentence contains some incoherence and occasional errors, but can still convey the general meaning to a certain extent.

4: The sentence is mostly fluent with only minor errors or slight awkwardness in expression, and effectively communicates the intended meaning.

5: Perfect. The sentence is completely fluent, free of any errors in grammar, punctuation, or expression, and clearly conveys the idea.

\item Emotion:

1: The emotional expression is extremely inappropriate and chaotic, not in line with the content, and may convey wrong emotions.

2: The emotional expression has obvious flaws, either too weak or exaggerated, and is disjointed from the content.

3: The emotional expression is average. It can convey basic emotions but lacks depth and has minor issues.

4: The emotional expression is good. It can effectively convey the intended emotion with an appropriate intensity and is well integrated with the content.

5: The emotional expression is excellent. It is rich, nuanced, and perfectly matches the content, capable of evoking a strong and appropriate emotional response.

\item Acceptance:

1: The response inescapably triggers emotional resistance.

2: The response is highly likely to trigger emotional resistance.

3: The response has a possibility of emotional resistance occurring.

4: The response rarely provokes emotional resistance.

5: The response has no occurrence of emotional resistance.

\item Effectiveness:

1:  The response actually worsens the seeker's emotional distress.

2: The response carries the risk of increasing stress levels, and this outcome varies depending on the individual user.

3: The response fails to alter the seeker's current emotional intensity and keeps it at the same level.

4: The response shows promise in calming the emotional intensity; however, it is overly complicated or ambiguous for the user to fully comprehend and utilize effectively.

5: The response appears to be highly effective in soothing the seeker's emotions and offers valuable and practical emotional support. 

\item Sensitivity:

1: The response renders inaccurate evaluations regarding the seeker's state.

2: The response is characterized by rash judgments, as it lacks adequate assessment and in-depth exploration of the seeker's state.

3: The response is formulated with a one-sided judgment and a limited exploration of the seeker's state.

4: The response demonstrates an understanding that only covers a part of the seeker's state.

5: The response precisely grasps the seeker's state and is appropriately tailored according to the seeker's actual situation.

\item Alignment:

1: The response is in total contradiction to the predicted strategy.

2: The response has a minor deviation from the predicted strategy.

3: There is some ambiguity between the response and the predicted strategy.

4: The response largely matches the predicted strategy, yet it contains some ambiguous elements.

5: The response effectively makes itself consistent with the predicted strategy.

\item Satisfaction:

1: The response is extremely disappointing. It doesn't answer the question at all and is of no help.

2: The response is poor. It only gives a partial answer and leaves many doubts unresolved.

3: The response is average. It meets the basic requirements but isn't particularly outstanding.

4: The response is good. It answers the question clearly and provides some useful details.

5: The response is excellent. It not only answers the question perfectly but also offers valuable additional insights.
\end{itemize}

\section{More Experimental Results}

\subsection{The Self-EmoQ Algorithm}
\label{Alg:EmoQ}

We illustrate the training mechanism of {\ModelName} by the pseudo-codes in Algorithm \ref{alg:emoq}.

\begin{algorithm}
\caption{Self-EmoQ}  
\begin{algorithmic}[1]
    \State Initialize the batch sizes $B$
    \State Initialize replay buffer $\mathcal{B}$ with original dataset, Q $Q_{\theta}(s, a)$, target Q $\hat{Q}_{\theta^-}(s, a)$ with $\theta^- \leftarrow \theta$
    \State Load pretrained model $g$, Set exploration rate $\epsilon$, discount factor $\gamma$, and update interval $C$, weight of Plutchik score $w$.
    
    \While{ not converged }
        \State Draw $B$ data $\{(h_t,x_t^u)\}$ from $\mathcal{B}$
    \For{each data}
        \State Form state $s_{t} =(h_t,s_t^u)$
        \State Select $e_{t}^s$ with $\epsilon$-greedy policy by Eq(\ref{eq:maxQ})
        \State Obtain $x_t^s$ , $r_t$ by Eq(\ref{eq:reward})
        \State Update history $h_{t+1}$;  state $s_{t+1}$ 
        \State Store transition $(s_t, a_t, r_t, s_{t+1})$ in $\mathcal{B}$
        \EndFor
    \State Sample batch $\{(s_i, a_i, r_i, s_i')\}$ from $\mathcal{B}$
    \State Compute TD target:
    $$y_i = r_i + \gamma \max_{a'} \hat{Q}_{\theta^-}(s_i', a')$$
    \State Update Q-network by minimizing Eq.(\ref{eq:dqn})
    \State Every {$C$} steps update $\hat{Q}$ by $\theta^- \leftarrow \theta$
    \EndWhile
\end{algorithmic}
\label{alg:emoq}
\end{algorithm}

\subsection{Inter-Annotator Consistence}
\label{inter-annotator agreement}

To indicate the scoring consistences of different human annotators, we analyze the their Fluency scoring statistics and present the Cohen’s kappa matrix in Table \ref{tab:Kappa_matrix}. It can be observed that most $\kappa$ values fall between 0.8 and 1, indicating high inter-annotator agreement.

\begin{table*}[!t]
    \centering
    \small
    \renewcommand{\arraystretch}{1.5}
\begin{tabular}{c| l}
        \Xhline{2\arrayrulewidth}
        \multicolumn{2}{c}{\textbf{Case of DailyDialog}}  \\ 
        \hline
        History & \makecell[l]{user: (\textit{no emotion}) Hi , Monica , how is everything going.\\
assistant: (\textit{no emotion}) Everything goes well , but I am thinking about quitting my current job.\\
user: (\textit{surprise}) Why. You ’ re not satisfied anymore.\\
assistant: (\textit{no emotion}) I just sense. But I cannot grow anymore. My boss is not really supporting me. \\I am interested in some positions in other JV companies , but I need to do some more in-depth research\\ before I send my application letters out.\\
user: (\textit{happiness}) That is important. Doing research on a company you are interested in will\\ definitely help your application.} \\
        \hline
        0-shot & \makecell[l]{(\textit{surprise})That's a good point, I didn't think about the importance of researching\\ the company before applying.}   \\
        \hline
        EmoFSM & \makecell[l]{(\textit{no emotion}) Thank you for your advice.}  \\ 
        \hline
        \textbf{\ModelName} & \makecell[l]{(\textit{happiness}) I will also need to update my resume. I've heard that the most important thing in a resume \\ is to make it short and concise.}    \\ 
        
        \Xhline{2\arrayrulewidth}
        \multicolumn{2}{c}{\textbf{Case of EmoryNlP}}  \\ 
        \hline
        History & \makecell[l]{Joey Tribbiani: (\textit{Neutral}) Hello.\\
Chandler Bing: (\textit{Scared}) Hey, it's me. I know you can't stand to be in the same room as me, so I just \\ thought I'd try and apologize over the phone. All I...\\
Joey Tribbiani: (\textit{Neutral}) Hello.\\
Chandler Bing: (\textit{Mad}) Look I never should have kissed your girlfriend, but I'm...} \\
        \hline
        0-shot & \makecell[l]{(\textit{Anger})Oh, so now you're apologizing? You think a simple phone call is going to make everything okay?\\ You kissed my girlfriend, Chandler. That's not something you just apologize for over the phone.}    \\
        \hline
        EmoFSM & \makecell[l]{(\textit{Neutral}) I don't want to listen to it.}   \\ 
        \hline
        \textbf{\ModelName} & \makecell[l]{(\textit{Anger}) I don't want to hear it. }     \\ 
        \Xhline{2\arrayrulewidth}
        \multicolumn{2}{c}{\textbf{Case of MELD}}  \\ 
                \hline
        History & \makecell[l]{Joey: (\textit{sadness}) God, it's gonna so weird like when I come home and you're not here.\\
        Joey: (\textit{sadness}) No more Joey and Chan's.\\
        Joey: (\textit{sadness}) No more J and C's.\\
        Joey: (\textit{neutral}) "You wanna go over to Joey and Chandler's?"\\
        Joey: (\textit{neutral}) "Can't, its not there."\\
        Chandler: (\textit{neutral}) Look, I'm just gonna be across the hall, we can still do all the same stuff
}  \\
        \hline
        0-shot & \makecell[l]{(\textit{Neutral}) Oh, okay, that sounds cool, we can still hang out.}   \\
        \hlineƒ
        EmoFSM & \makecell[l]{(\textit{sadness}) No we can't!}    \\ 
        \hline
        \textbf{\ModelName} & \makecell[l]{(\textit{sadness}) Yeah, but it's not gonna be the same.}     \\ 
        \Xhline{2\arrayrulewidth}
        
    \end{tabular}

    \caption{More Good Cases of {\ModelName}.} 
    \label{tab:more_good_cases}
\end{table*}

\subsection{More Typical Cases}
\label{more case}


Besides the good case of IEMOCAP shown in Table \ref{tab:good_cases}, Table \ref{tab:more_good_cases} further presents representative cases from DailyDialog, EmoryNLP, and MELD. These examples consistently demonstrate that \ModelName can generate responses that are not only coherent with the dialogue context but also better aligned with the underlying emotional dynamics. In the DailyDialog case, \ModelName produces a natural continuation that maintains the positive tone of the conversation while further developing the ongoing topic of job application preparation. In the EmoryNLP case, it appropriately captures Joey's anger toward Chandler and generates a concise response that is consistent with both the preceding conflict and the target emotion. Similarly, in the MELD case, \ModelName recognizes Joey's sadness about the upcoming change in his relationship with Chandler and responds with an emotionally appropriate continuation that preserves the interpersonal context. Compared with the 0-shot and EmoFSM baselines, these cases illustrate that \ModelName achieves a better balance between contextual coherence and emotion-consistent response generation across different dialogue domains.

\subsection{Human Verification on Plutchik Scores} \label{Evaluation of GPT Scoring}

To further validate the reliability of our GPT-based rewarding mechanism (the Plutchik score $r_{\text{Plu}}$), we conduct additional human annotations on these rewards to validate their consistence. To achieve this, we randomly select 100 samples, then ask the volunteers to annotate the reward according to the \textbf{Plutchik’s Wheel of Emotion} theory, on the same reward scale and dimensions: Alignment, Transition, and Function. We report means and standard deviations of human and GPT-4o scores, as well as the correlation coefficient $\rho$ and Cohen's kappa $\kappa$ between them, to indicate their correlation levels.

Table \ref{tab:comparison_plu} exhibits this experiment's results. While the GPT-4o scores differ slightly from the human scores in terms of mean values, the results of $\rho$ and $\kappa$ indicate a high level of agreement between these two rewarding methods.

\begin{table}[!ht]
    \centering
    \resizebox{0.7\columnwidth}{!}{
    \begin{tabular}{lllll}
    \toprule
        ~ & mean & std & $\rho$ & $\kappa$  \\ 
    \midrule
     Alignment: &~ &  ~ & ~ & ~  \\ 
    \midrule
        GPT-4o & 4.31 & 1.00 & 0.87 & 0.58 \\ 
        Human & 4.16 & 0.94 & - & - \\ 
    \midrule
    Emotion: &~ &  ~ & ~ & ~ \\ 
    \midrule
        GPT-4o & 4.13 & 1.16 & 0.96 & 0.78 \\ 
        Human & 4.26 & 1.08 & - & - \\ 
    \midrule
    Effectiveness: &~ &  ~ & ~ & ~ \\ 
    \midrule
        GPT-4o & 3.86 & 1.33 & 0.95 & 0.72 \\ 
        Human & 3.69 & 1.21 & - & - \\ 
    \bottomrule
    \end{tabular}}
\caption{Statistical correlations of $r_{Plu}$ between human and GPT-4o.}
\label{tab:comparison_plu}
\end{table}

\subsection{Sensitivity Study on $w$}
\label{sensitivity}

The weight of Plutchik score, $w$ is an important hyper-parameter, where higher values of $w$ emphasize theory-driven emotional consistency, whereas lower values rely more heavily on dataset annotations. Therefore, we conduct a sensitivity study on $w$, with results shown in Table~\ref{tab:reward_w}. Results show that our framework is flexible and robust to different settings of $w$, while the optimal balance differs across datasets. This dataset-dependent characteristic suggests that different corpora exhibit varying degrees of annotation noise and emotional ambiguity, and our reward formulation allows this trade-off to be adjusted accordingly.

\begin{table}[htbp!]
\centering
\renewcommand{\arraystretch}{1.1}
\fontsize{14pt}{16pt}\selectfont
\resizebox{0.98\columnwidth}{!}{
\begin{tabular}{llllllll}
\toprule
$w$ & $1$ & $0.9$ &  $0.8$ & $0.7$ &  $0.5$ &  $0.3$ & $0$ \\ 
\midrule
DailyDialog & 0.78 & 0.73 & 0.69 & 0.65 & 0.57 & 0.48 & 0.35 \\
EmoryNLP      &0.53 & 0.58 & 0.61 & 0.65 & 0.71 & 0.79 & 0.88 \\
MELD          &0.82 & 0.82 & 0.81 & 0.81 & 0.86 & 0.91 & 0.85 \\
IMEOCAP       &0.53 & 0.58 & 0.61 & 0.67 & 0.75 & 0.83 & 0.88\\
\bottomrule
\end{tabular}}
\caption{Averaged rewards obtained on different values of $w$.}
\label{tab:reward_w}
\end{table}



\end{document}